\documentclass[twocolumn]{revtex4}
 
\usepackage{graphicx}
\usepackage{color}

\newcommand{\RefTitle}[1]{#1,}

\usepackage{ulem}

\begin{document}
\title{Increasing decoherence rate of Rydberg polaritons due to accumulating dark Rydberg atoms
}

\author{Ko-Tang Chen,$^1$
Bongjune Kim,$^{1,}$\footnote{Electronic address: {\tt upfe11@gmail.com}}
Chia-Chen Su,$^1$
Shih-Si Hsiao,$^1$
Shou-Jou Huang,$^2$
Wen-Te Liao,$^{2,3,4}$
and
Ite A. Yu,$^{1,4,}$
}
\email{yu@phys.nthu.edu.tw}

\address{$^1$Department of Physics, National Tsing Hua University, Hsinchu 30013, Taiwan \\
$^2$Department of Physics, National Central University, Taoyuan City 320317, Taiwan \\
$^3$Physics Division, National Center for Theoretical Sciences, Taipei 10617, Taiwan\\ 
$^4$Center for Quantum Technology, Hsinchu 30013, Taiwan
} 

\begin{abstract}
We experimentally observed an accumulative type of nonlinear attenuation and distortion of slow light, i.e., Rydberg polaritons, with the Rydberg state $|32D_{5/2}\rangle$ in the weak-interaction regime. The present effect of attenuation and distortion cannot be explained by considering only the dipole-dipole interaction (DDI) between Rydberg atoms in $|32D_{5/2}\rangle$. Our observation can be attributed to the atoms in the dark Rydberg states other than those in the bright Rydberg state, i.e., $|32D_{5/2}\rangle$, driven by the coupling field. The dark Rydberg states are all the possible states, in which the population decaying from $|32D_{5/2}\rangle$ accumulated over time, and they were not driven by the coupling field. Consequently, the DDI between the dark and bright Rydberg atoms increased the decoherence rate of the Rydberg polaritons. We performed three different experiments to verify the above hypothesis, to confirm the existence of the dark Rydberg states, and to measure the decay rate from the bright to dark Rydberg states. In the theoretical model, we included the decay process from the bright to dark Rydberg states and the DDI effect induced by both the bright and dark Rydberg atoms. All the experimental data of slow light taken at various probe Rabi frequencies were in good agreement with the theoretical predictions based on the model. This study pointed out an additional decoherence rate in the Rydberg-EIT effect, and provides a better understanding of the Rydberg-polariton system.
\end{abstract}

\maketitle

\newcommand{\FigOne}{
	\begin{figure}[t]
	\center{\includegraphics[width=80mm]{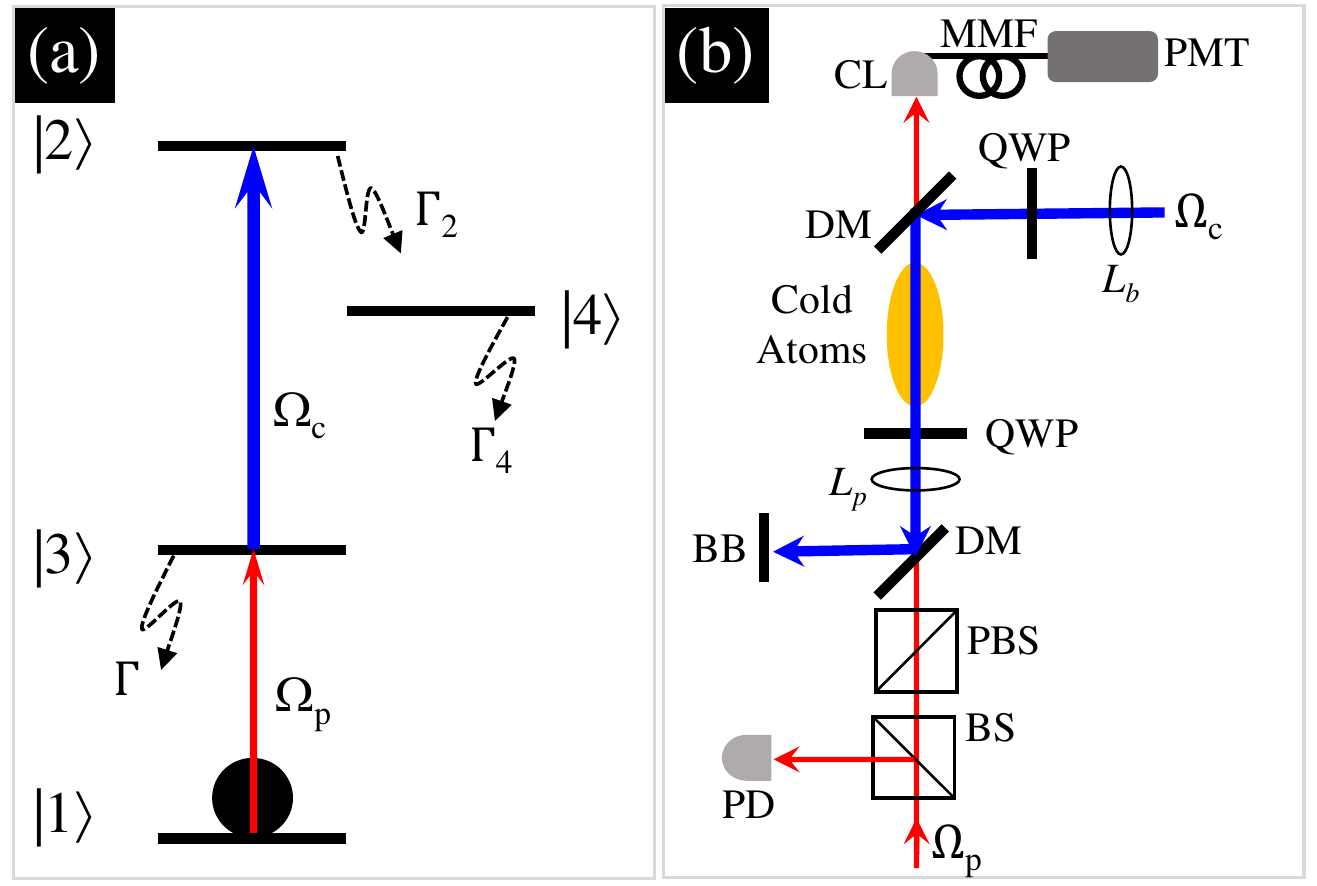}}
	\caption{
(a) Relevant energy levels and transitions of the laser fields. $|1\rangle$ and $|3\rangle$ are the ground and intermediate excited states. $|2\rangle$ is the Rydberg state driven by the coupling field, and $|4\rangle$ represents all the nearby Rydberg states of $|2\rangle$ that involve the DDI effect. $|2\rangle$ and $|4\rangle$ are the bright and dark Rydberg states, respectively. $\Gamma$, $\Gamma_2$, and $\Gamma_4$ represent the spontaneous decay rates of $|3\rangle$, $|2\rangle$, and $|4\rangle$. $\Omega_{p}$ and $\Omega_c$ denote the Rabi frequencies of the probe and coupling fields. The two fields' frequencies were stabilized to maintain the two-photon resonance. As compared with $\Gamma$, the magnitude of the one-photon detuning was negligible. (b) Sketch of the experiment setup. BS: beam splitter; PBS: polarizing beam splitter; PD: photo detector; DM: dichroic mirror; $L_p$, $L_b$: lenses; QWP: quarter-wave plate; CL: collimation lens; MMF: multimode optical fiber; PMT: photomultiplier tube. Red and blue arrowed lines indicate the optical paths of the probe and coupling beams.
	}
	\label{fig:EIT_transition}
	\end{figure}
}
\newcommand{\FigTwo}{
	\begin{figure}[t]
	\center{\includegraphics[width=55mm]{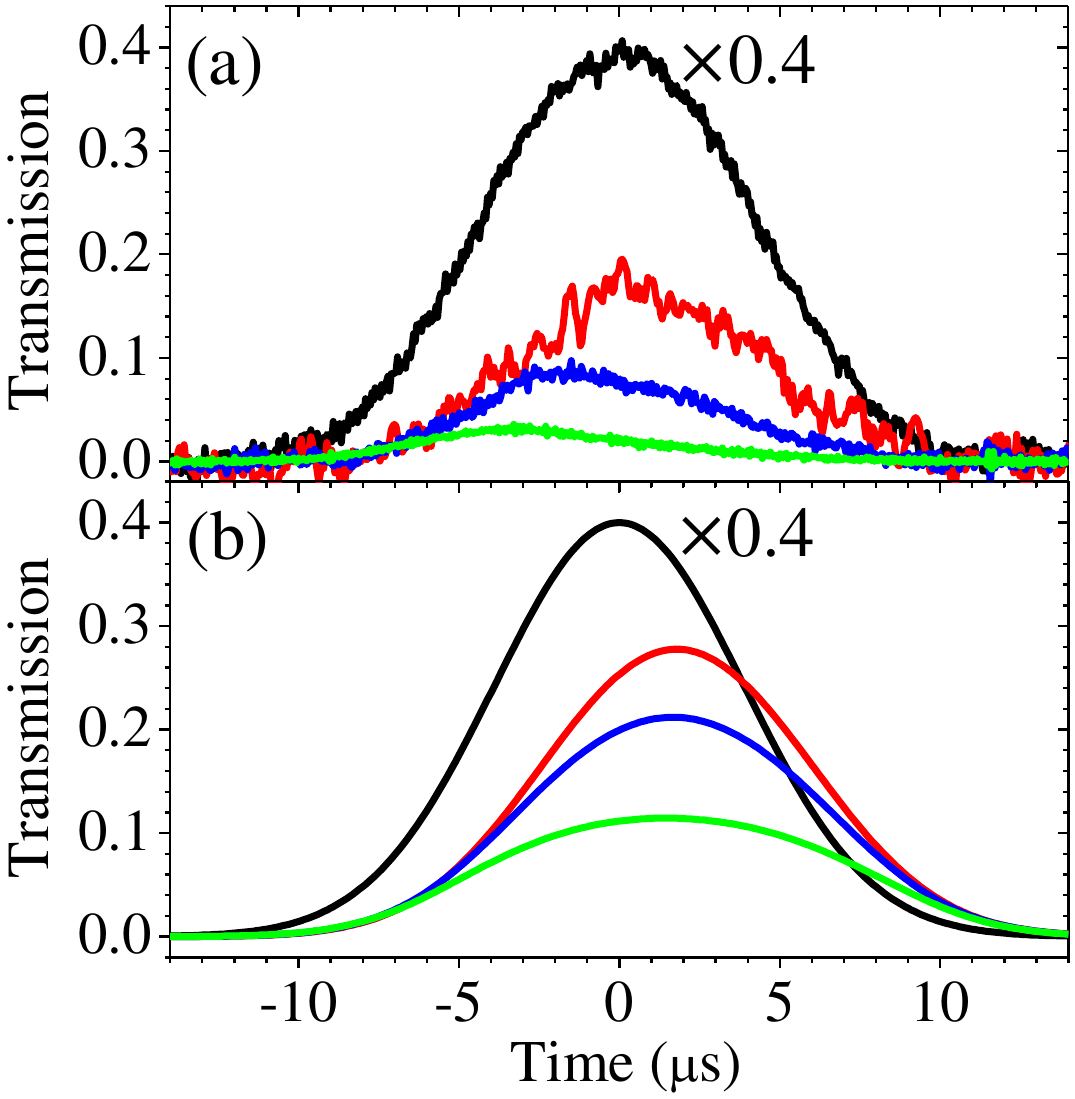}}
	\caption{Experimental observation that the slow light of a Gaussian pulse was distorted and a larger value of $\Omega_{p0}$ made the distortion more severe. (a) Experimental data of the output pulses versus time are shown by red, blue, and green lines, and their input Rabi frequencies, $\Omega_{p0}$, of 0.05$\Gamma$, 0.1$\Gamma$, and 0.2$\Gamma$, respectively. Since the shape of the three input pulses is very similar, we only plot the one with the $\Omega_{p0}$ of 0.1$\Gamma$ as shown by the black line, which is scaled down by a factor of 0.4. In the measurements, $\alpha$ (optical depth) = 70, $\Omega_c =$ 1.0$\Gamma$, and $\gamma_0 =$ 9.0$\times10^{-3}\Gamma$, which were determined experimentally \cite{OurPRA2019}. (b) Theoretical predictions by solving Eqs.~(\ref{eq:OBE_first})-(\ref{eq:OBE_last}) with the above values of $\Omega_{p0}$, $\alpha$, $\Omega_c$, and $\gamma_0$, and with the coefficient $A$ = 0.76$\Gamma$ given by Eq.~(\ref{definitionA}).
}
	\label{fig:Gaussian1}
	\end{figure}
}

\newcommand{\FigThree}{
	\begin{figure}[t]
	\center{\includegraphics[width=87mm]{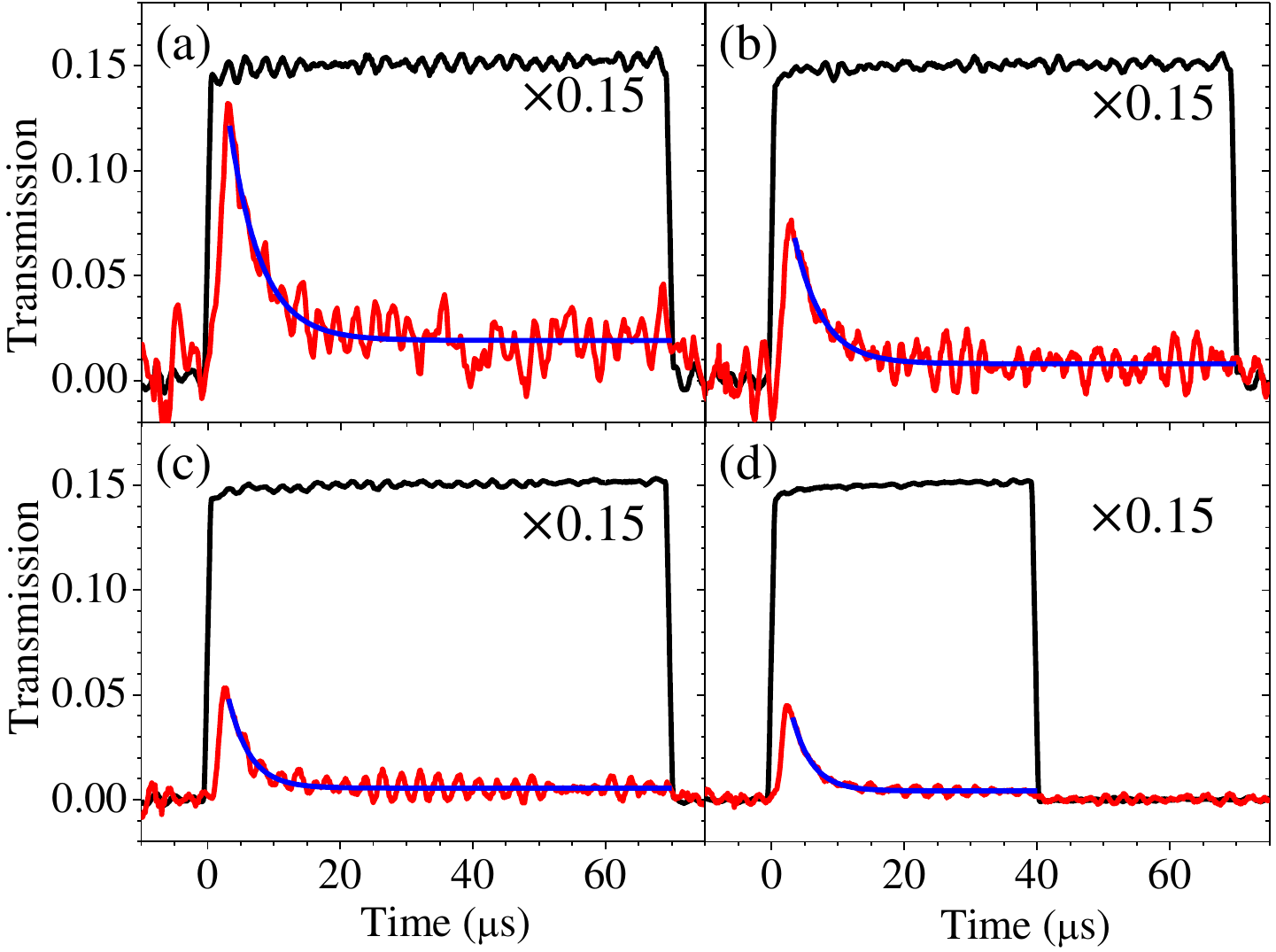}}
	\caption{Experimental observation of the accumulation effect, in which the DDI-induced attenuation increased with time. In each panel, the black line scaled down by a factor of 0.15 is the input pulse, the red line represents the output pulse, and the blue line is the best fit of the tail part of the red line. The best fit is an exponential-decay function. The values of $\Omega_{p0}$ were (a) 0.08$\Gamma$, (b) 0.1$\Gamma$, (c) 0.15$\Gamma$, and (d) 0.2$\Gamma$, respectively. In the measurements, $\alpha$ (optical depth) = 70, $\Omega_c =$ 1.0$\Gamma$, and $\gamma_0 =$ 1.1$\times10^{-2}$$\Gamma$, which were determined experimentally \cite{OurPRA2019}.
	}
	\label{fig:SquarePulse1}
	\end{figure}
}
\newcommand{\FigFour}{
	\begin{figure}[t]
	\center{\includegraphics[width=56mm]{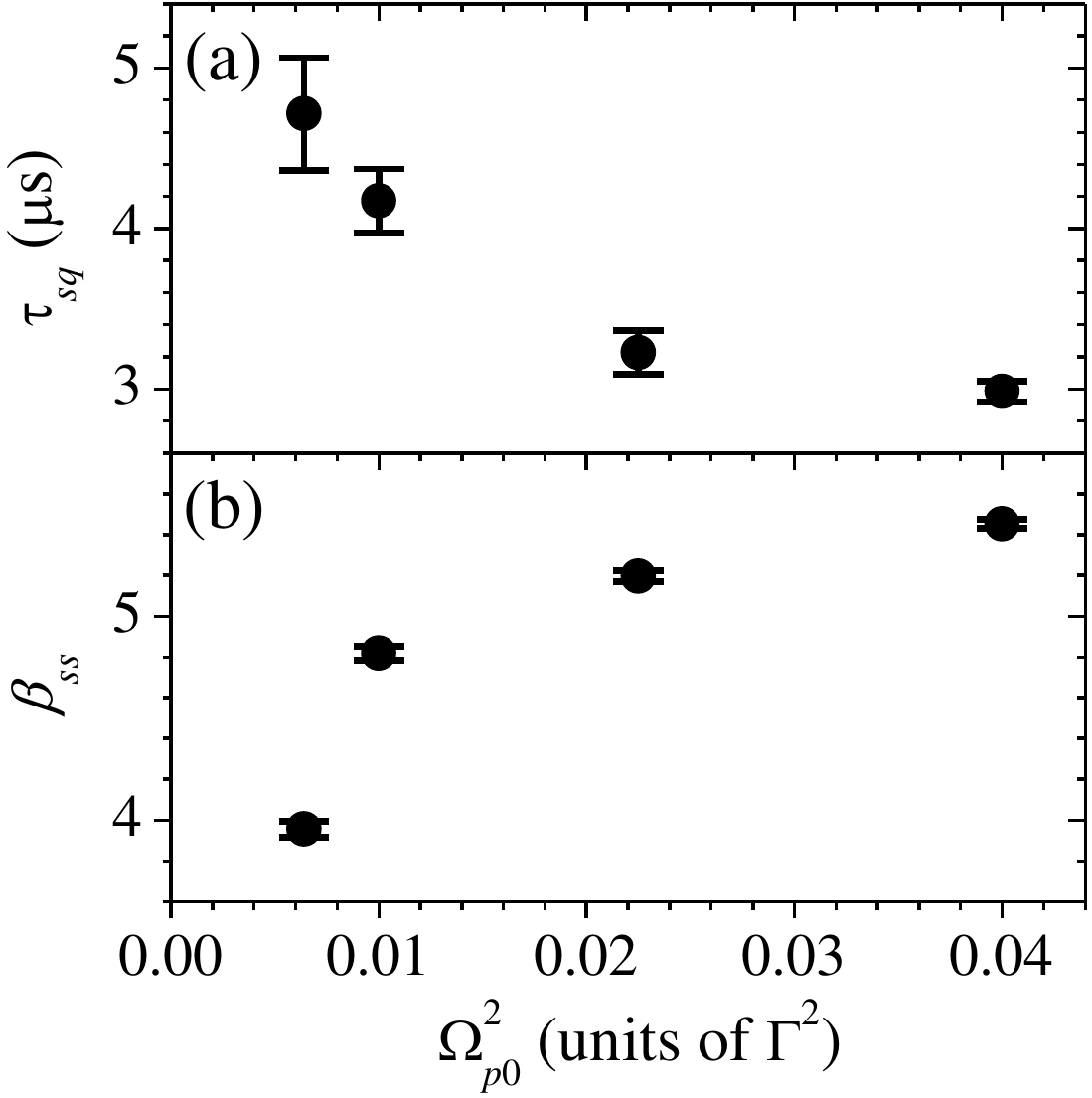}}
	\caption{Experimental observation of a larger $\Omega_{p0}$ resulting in a faster accumulation of DDI-induced attenuation. (a) Decay time constant, $\tau_{sq}$, as a function of $\Omega_{p0}^2$. The values of $\tau_{sq}$ were determined by the best fits shown in Fig.~\ref{fig:SquarePulse1}. (b) The steady-state attenuation coefficient, $\beta_{ss}$, as a function of $\Omega_{p0}^2$. $\beta_{ss} = -\ln{(T_{ss})}$, where $T_{ss}$ is the steady-state transmission. The values of $T_{ss}$ were determined by the vertical offsets of the blue lines shown in Fig.~\ref{fig:SquarePulse1}.
}
	\label{fig:DecayConstant}
	\end{figure}
}
\newcommand{\FigFive}{
	\begin{figure}[t]
	\center{\includegraphics[width=70mm]{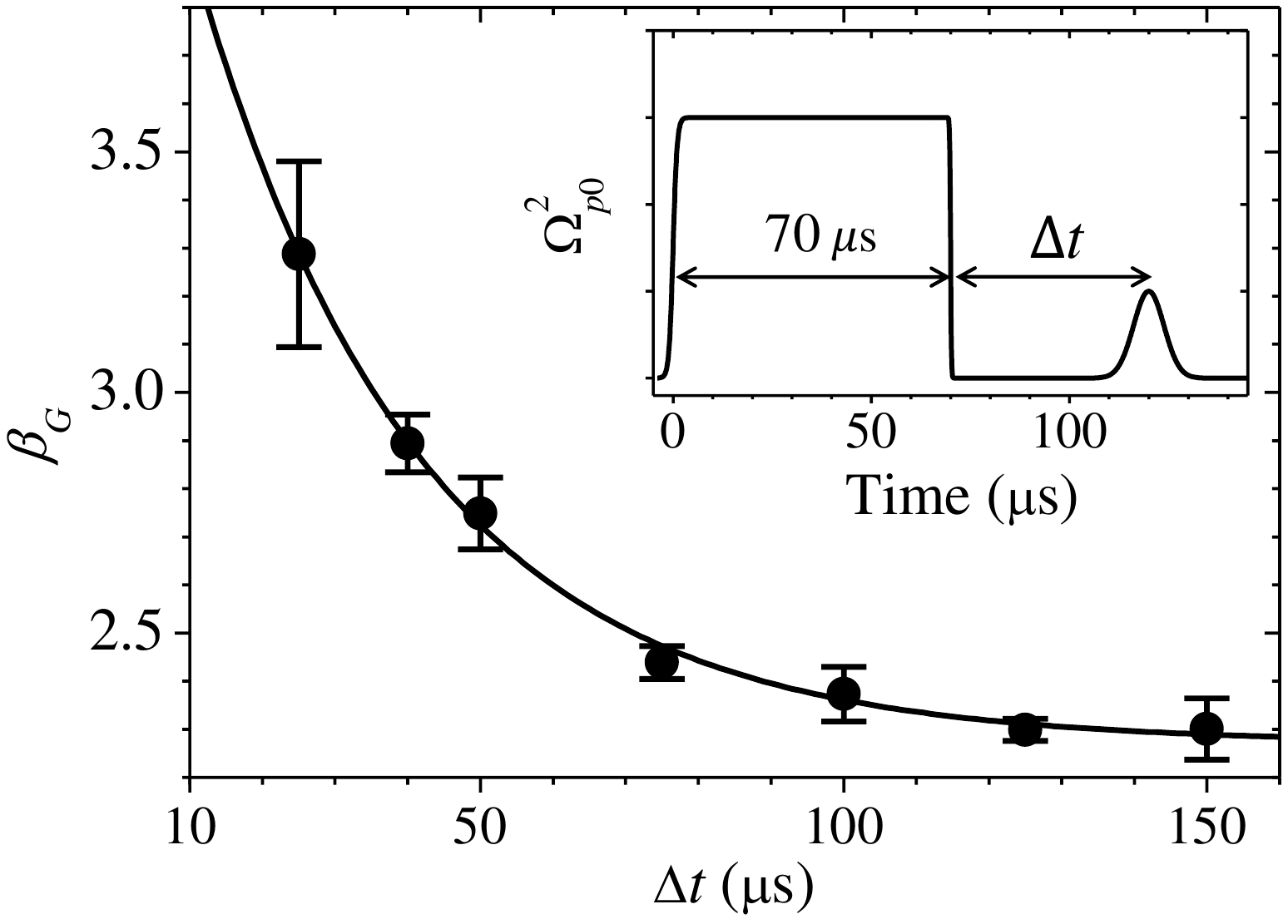}}
	\caption{Experimental observation of the lifetime of the dark Rydberg states by the measurement of DDI-induced attenuation $\beta_G$ as a function of $\Delta t$. Inset: the time sequence of the measurement. The coupling field was continuously on during the measurement. Circles are the experimental data and the black line is the best fit of an exponential-decay function. The decay time constant of the best fit was 31 $\mu$s, which is similar to the lifetimes of dark Rydberg states with the principal quantum number, $n$, of around 32.
}
	\label{fig:PollutantDecayTime}
	\end{figure}
}
\newcommand{\FigSix}{
	\begin{figure}[t]
	\center{\includegraphics[width=61mm]{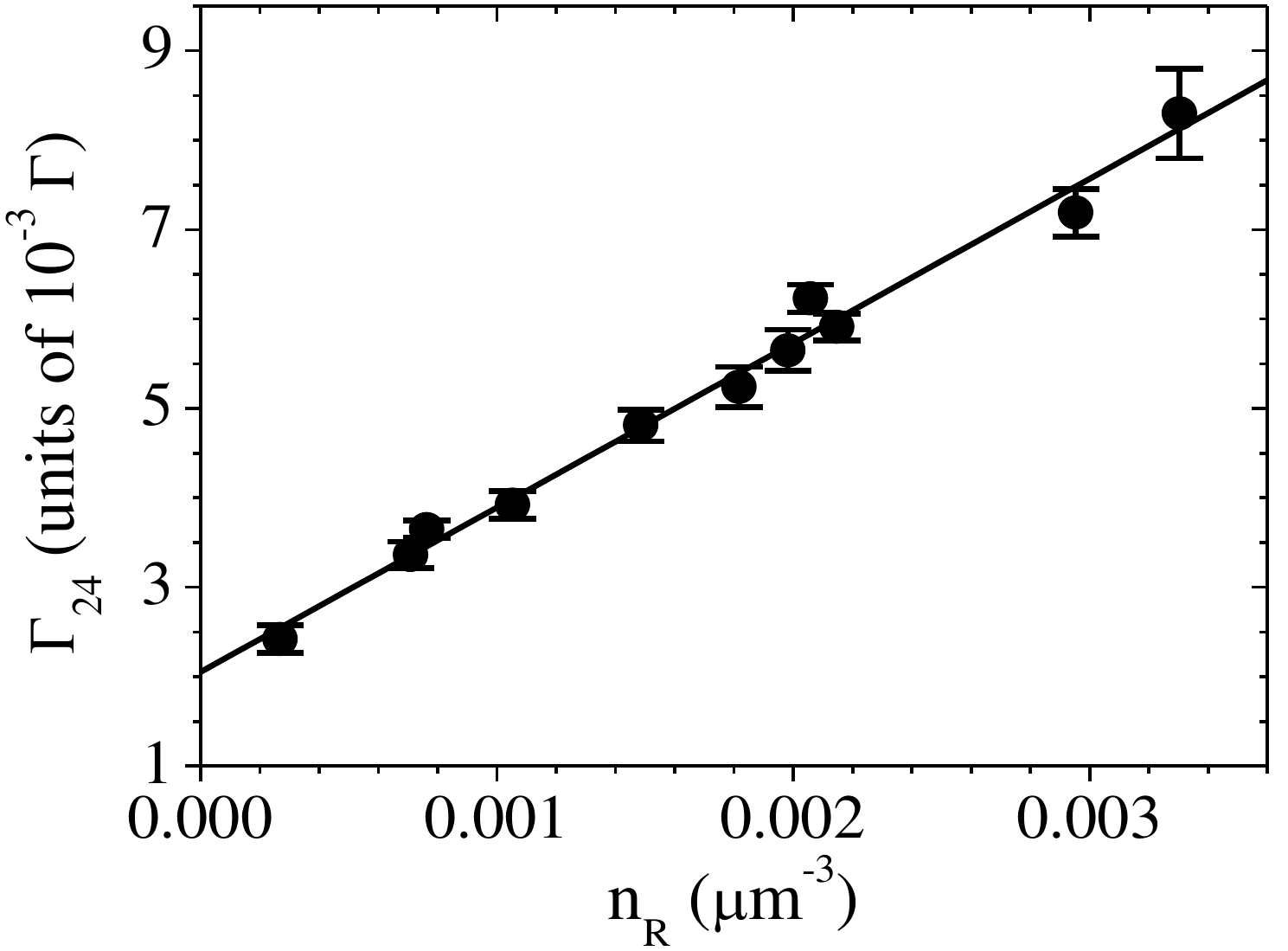}}
	\caption{The decay rate $\Gamma_{24}$ as a function of the density, $n_R$, of the atoms in $|2\rangle$. Circles are the experimental data. Each data point was determined by the result of a series of measurements similar to those in Supplemental Material. The black line is the best fit of a straight line.
	}
	\label{fig:Gamma24}
	\end{figure}
}
\newcommand{\FigSeven}{
	\begin{figure}[t]
	\center{\includegraphics[width=87mm]{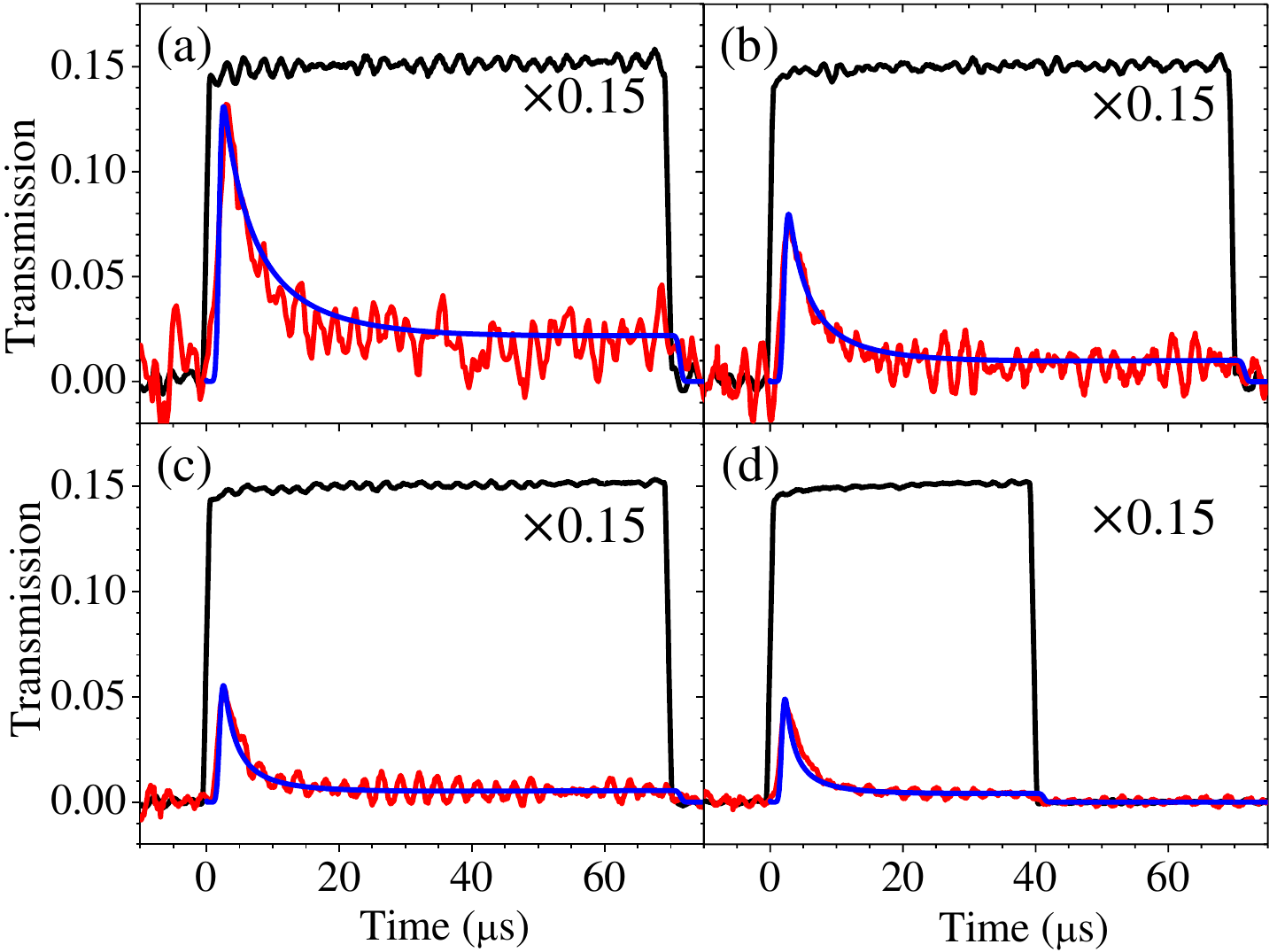}}
	\caption{Comparison between the experimental data of the long square probe pulses and the theoretical predictions. Black lines represent the input probe pulses and are scaled down by a factor of 0.15. Red lines represent the output probe pulses under the constant presence of the coupling field. The experimental data are the same as those shown in Fig.~\ref{fig:SquarePulse1}, where the input Rabi frequencies $\Omega_{p0}$ are (a) 0.08$\Gamma$, (b) 0.1$\Gamma$, (c) 0.15$\Gamma$, and (d) 0.2$\Gamma$. Blue lines are the predictions calculated using Eqs.~(\ref{eq:MSE}) and (\ref{eq:mOBE_first})-(\ref{eq:mOBE_last}). In the calculation, we set $\alpha$ (optical depth) = 70, $\Omega_c =$ 1.0$\Gamma$, and $\gamma_0 =$ 1.1$\times$$10^{-2}$$\Gamma$, which were determined experimentally, the coefficients $C =$ $9.1$$\times$$10^{-2}$$\Gamma$ and $D =$ $1.3$$\times$$10^{-3}$$\Gamma$, which were deduced from the result of Fig.~\ref{fig:Gamma24}, and $A =$ 0.76$\Gamma$ given by Eq.~(\ref{definitionA}). $B$ was adjusted to make the predictions fit the data and its optimum value is 7.7$~\Gamma$.
	}
	\label{fig:SquarePulse2}
	\end{figure}
}
\newcommand{\FigEight}{
	\begin{figure}[t]
	\center{\includegraphics[width=55mm]{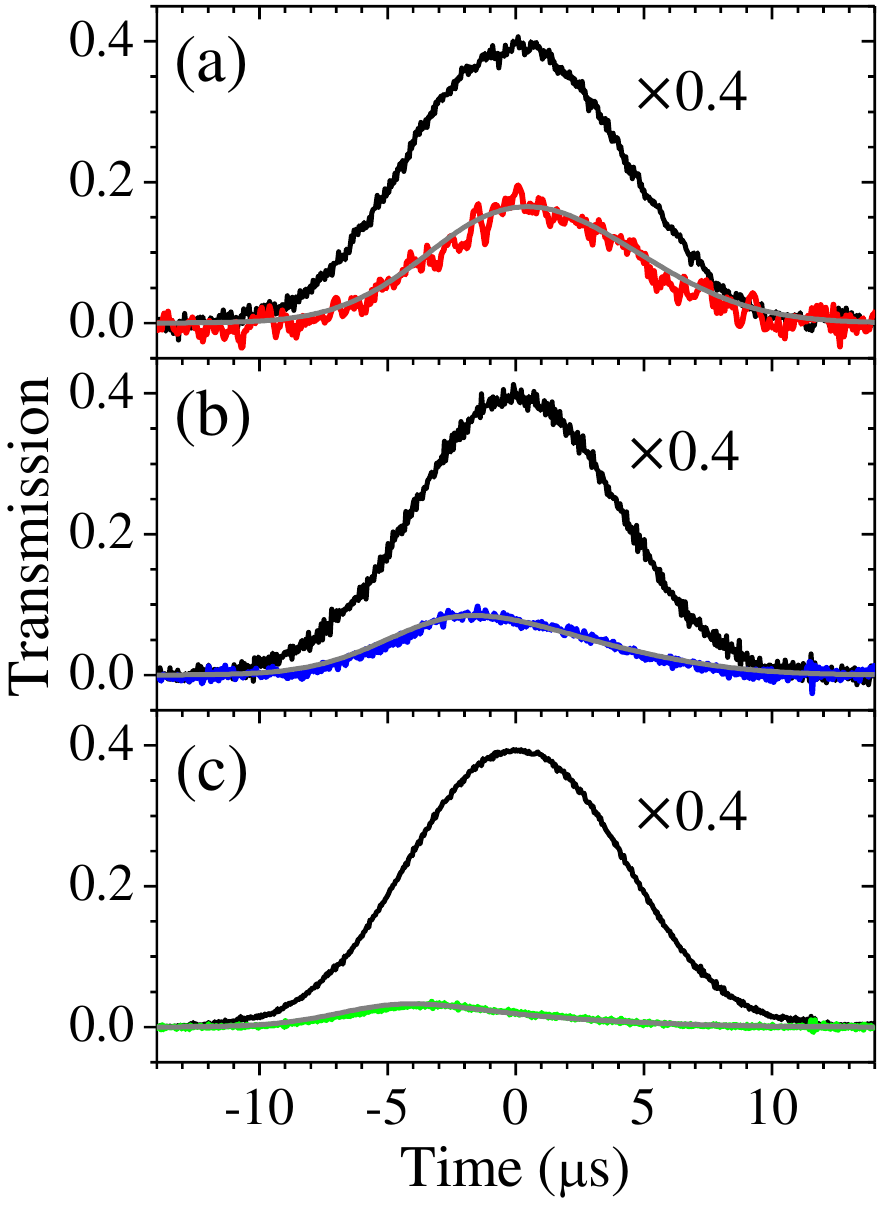}}
	\caption{Comparison between the experimental data of the Gaussian input probe pulses and the theoretical predictions. Black lines represent the input probe pulses and are scaled down by a factor of 0.4. Red, blue, and green lines represent the output probe pulses under the constant presence of the coupling field. The experimental data are the same as those shown in Fig.~\ref{fig:Gaussian1}, where the input Rabi frequencies $\Omega_{p0}$ are (a) 0.05$\Gamma$, (b) 0.1$\Gamma$, and (c) 0.2$\Gamma$. Gray lines are the theoretical predictions calculated with Eqs.~(\ref{eq:MSE}) and (\ref{eq:mOBE_first})-(\ref{eq:mOBE_last}). All the parameters used in the calculation here are the same as those used in Fig.~\ref{fig:SquarePulse2}, except for $\gamma_0 =$ 9.0$\times$$10^{-3}$$\Gamma$.
	}
	\label{fig:Gaussian2}
	\end{figure}
}
\section{Introduction}

Atoms in the Rydberg states possess a strong electric dipole-dipole interaction (DDI) among themselves. The blockade effect arising from the DDI is a versatile mechanism for quantum information processing \cite{blockade_Zoller2000, Comparat_2010, SaffmanRMP, Pfau_2013}. The combination of the DDI and the effect of electromagnetically induced transparency (EIT) can mediate strong photon-photon interactions \cite{Adams_2010, Fleischhauer_2011, Lukin_Nature_2012, Gorshkov_PRL_2017, Hofferberth_2014, Rempe_swithch_2014, Rempe_2014, Rempe_gate}. Thus, Rydberg atoms have led to applications such as quantum logic gates \cite{Lukin_2000, Saffman_2017, Lukin_PRL_2019, Rempe_gate}, single-photon generation \cite{Kuzmich_2012, Pfau_2018, ondemand_2020}, single-photon transistors \cite{Hofferberth_2014, Rempe_2014}, and single-photon switches \cite{Rempe_swithch_2014}. Furthermore, the system of Rydberg atoms in a vapor or an array is also a platform for the study of many-body physics \cite{Bloch_2012, dressing_2016, Gross_2016, Browaeys_2016, Lukin_2017, Lukin_2019, dressing_2020, Lukin_2021}.

Most of the studies on Rydberg atoms have been performed in a strong interaction regime. On the contrary, the DDI-induced effect in the weak interaction regime based on EIT was studied theoretically~\cite{OurOE2020} and experimentally~\cite{OurCommunPhys2021} with a low principal quantum number of $n = 32$ and a low Rydberg-atom density equal to or less than 2.0$\times$10$^9$~cm$^{-3}$. The weak interaction regime in those studies were $(r_B/r_a)^3 < 0.06$ where $r_a$ and $r_b$ are the half mean distance between the Rydberg atoms and the blockade radius, respectively. Due to the EIT effect \cite{EIT_review}, the propagation delay time or photon-matter interaction time in a high-OD medium was a couple of $\mu$s. With help of a sufficient collision rate and interaction time in a high-OD EIT medium of weakly interacting Rydberg atoms, it was observed that a smaller width of the transverse momentum distribution of Rydberg polaritons at the exit of the system as compared with that at the entrance~\cite{OurCommunPhys2021}.

The observation in Ref.~\cite{OurCommunPhys2021} is a demonstration of thermalization process of Rydberg polariton, and indicates a possibility of Rydberg polariton Bose-Einstein condensation (BEC). It was also suggested to make Rydberg polaritons stationary and cool down the temperature in both transverse and longitudinal directions for the realization of BEC. To achieve the Rydberg polariton BEC, it is needed to gather rather many particles of weakly-interacting polaritons with a long interaction time in the system. However, the loss of the polaritons due to an extra attenuation, which is discussed in this study, hinders the particles from achieving the Rydberg polariton BEC.

In this work, we systematically studied the transmission of a probe field propagating through a Rydberg-EIT system with experimental conditions similar to Ref.~\cite{OurCommunPhys2021}. We observed the distorted output pulses which were inconsistent with the theoretical prediction considering the DDI-induced decoherence rate~\cite{OurOE2020, OurCommunPhys2021}. Based on the measurements, we found evidence of the presence of the dark Rydberg states which were not driven by the coupling field. The population can be transferred from a bright Rydberg state, which was excited by coupling field, to the dark Rydberg states. Since not interacting with the coupling field, the dark Rydberg atoms accumulated over time. On the other hand, due to the DDI between the dark and bright Rydberg atoms, the DDI-induced decoherence rate increased with time. 

Following the scenario described in the previous paragraph, we proposed a theoretical model based on the experimental data. Using a theoretical model to take into account the effect that population accumulates in some dark Rydberg states and the dark Rydberg atoms cause an extra DDI-induced decoherence rate, we can successfully explain the experimental data. We believe the effect is universal. There are several previous papers that studied decay or transition from the bright to dark Rydberg states, and the phenomena caused by the dark Rydberg atoms. However, to our knowledge, the increasing DDI-induced decoherence rate due to atoms accumulating in dark Rydberg states is a new phenomenon, which has not been systematically studied before. This work can provide a better understanding for the creation of dark Rydberg atoms and their influence to Rydberg polaritons in the Rydberg-EIT system. This work also points out an obstacle in the realization of the BEC with weakly-interacting Rydberg polaritons. 

This article is organized as follows. In Sec.~\ref{sec:accumulation}, we present the distorted output probe pulse shapes, which we cannot explain with the theoretical model as it only considered the DDI between the bright Rydberg states under the EIT condition. In addition, we show the necessity of the consideration of the dark Rydberg states by measurements of the decay time constant, the steady-state attenuation coefficient of the square input pulse, and the lifetime of the dark Rydberg state. In Sec.~\ref{sec:modifiedtheory}, we introduce a modified theoretical model to describe the increasing attenuation coefficient over time depending on the density of the Rydberg atom. We also report the method for estimating the decay rate from the bright Rydberg state to the dark Rydberg state and compare the experimental data with the theoretical prediction. In Sec.~\ref{sec:discussion}, we discuss the possible transfer process from the bright to the dark Rydberg states and the interaction strength between a bright and a dark Rydberg atoms. Finally, we summarize the results in Sec.~\ref{sec:conclusion}.

\section{Observation of the Accumulation Effect}
\label{sec:accumulation}

\subsection{Experiment setup}

The experiment was performed with a cigar-shaped cold $^{87}$Rb atom cloud produced by a magneto-optical trap. The dimension of the cloud was $1.5\times1.5\times5.0$~mm$^3$~\cite{cigar-shape MOT} and the temperature was about 350~$\mu$k~\cite{OurPRA2019,OurCommunPhys2021}. The transition scheme formed by the probe and coupling fields is shown in Fig.~\ref{fig:EIT_transition}(a). The probe field drove the transition between the ground state $|1\rangle$ and the intermediate state $|3\rangle$, and the coupling field drove the transition between state $|3\rangle$ and the Rydberg state $|2\rangle$. States $|1\rangle$, $|2\rangle$, and $|3\rangle$ corresponded to the ground state $|5S_{1/2}, F=2, m_F=2\rangle$, the Rydberg state $|32D_{5/2}, m_J=5/2\rangle$, and the excited state $|5P_{3/2}, F=3, m_F=3\rangle$ of $^{87}$Rb atoms, respectively. We prepared all population in a single Zeeman state $|5S_{1/2}, F=2, m_F=2\rangle$ by optical pumping~\cite{Natcomm_2014}. In the experiment, the $\sigma_+$-polarized probe and coupling fields were used. Owing to the optical pumping and polarization of the laser fields, the relevant energy levels $|1\rangle$, $|2\rangle$, and $|3\rangle$ were considered as a single Zeeman state. The spontaneous decay rate of $|3\rangle$ is $\Gamma=2\pi \times$6.07~MHz, and that of $|2\rangle$ is $\Gamma_2 =$ 2$\pi$$\times$7.9 kHz or 1.3$\times$$10^{-3}$$\Gamma$~\cite{Rb87Data, RydbergStateLifeTime}. As the van der Waals interaction energy is denoted as $\hbar C_6/r^6$, the Rydberg atoms in $|32D_{5/2}, m_J = 5/2\rangle$ have $C_6 =-2\pi\times$130~MHz$\cdot\mu$m$^6$~\cite{Saffman_C6}.

\FigOne

The experimental scheme is shown in Fig.~\ref{fig:EIT_transition}(b). Frequency stabilized laser systems generated the probe and coupling fields. The details of the stabilization method are described in Refs.~\cite{OurPRA2019,OurCommunPhys2021}. The probe and coupling fields were the first-order beams of the acousto-optic modulators (AOMs, not drawn in the figure). The AOMs were used to control the time sequence, shape the probe pulse, and adjust the frequency and the amplitude of the fields precisely. Each field was sent to the atom cloud by the polarization maintained optical fiber (PMF) after the AOM. To minimize the Doppler effect, we used the counter-propagating scheme of the probe and coupling fields. At the center of the cloud, the $e^{-1}$ full widths of the probe and coupling beams were 130 and 250~$\mu$m, respectively. The pulse shape and amplitude of the input probe field were monitored by a photo detector (PD) before entering the atom cloud as shown in Fig.~\ref{fig:EIT_transition}(b). The probe field was detected by a photomultiplier tube (PMT) after passing through the atom cloud. A digital oscilloscope (Agilent MSO6014A) acquired the signal from the PMT and produced the raw data.

\subsection{Theoretical model without the accumulation effect}
\label{subsec:theorywoaccumulation}

The Rydberg atoms inside the cloud were considered randomly distributed particles, similar to ideal gases, due to the assumption that the condition of a weakly-interacting many-body system of Rydberg polaritons was satisfied. Therefore, the mean-field model developed in Ref.~\cite{OurOE2020} was adopted to describe the DDI-induced behaviors in this study. 

We initially considered only the states $|1\rangle$, $|2\rangle$, and $|3\rangle$ shown by Fig.~\ref{fig:EIT_transition}(a) in the theoretical model. The DDI-induced decoherence rate ($\gamma_{\rm DDI}$) and frequency shift ($\delta_{\rm DDI}$), which are caused by the population in the Rydberg state $|2\rangle$, were taken into account. The optical Bloch equations (OBEs) of the density matrix operator and the Maxwell-Schr\"{o}dinger equation (MSE) of the probe field are given below: 
\begin{eqnarray}
\label{eq:OBE_first}
\label{eq:OBE_rho21}
	\frac{\partial}{\partial t}\rho_{21} &=& 
		\frac{i}{2}\Omega_{c}\rho_{31} +i (\delta_{\rm DDI} + \delta) \rho_{21}
		\nonumber \\
		&-& \left( \gamma_{\rm DDI} + \gamma_0+\frac{\Gamma_2}{2} \right) \rho_{21}, \\
\label{eq:OBE_rho31}		
	\frac{\partial}{\partial t}\rho_{31} &=&
		\frac{i}{2}\Omega_{p}(\rho_{11}-\rho_{33})
		+\frac{i}{2}\Omega_{c}\rho_{21}+ i\Delta_{p}\rho_{31} \nonumber \\
		&-&\frac{\Gamma}{2}\rho_{31}, \\
\label{eq:OBE_rho32}
	\frac{\partial}{\partial t}\rho_{32} &=&
		\frac{i}{2}\Omega_{p}\rho_{21}^* +\frac{i}{2}\Omega_{c}(\rho_{22}-\rho_{33}) \nonumber \\
		&-&i \Delta_c \rho_{32}-\frac{\Gamma}{2} \rho_{32}, \\
\label{eq:OBE_rho22}
	\frac{\partial}{\partial t}\rho_{22} &=&
		\frac{i}{2}\Omega_{c}\rho_{32}-\frac{i}{2}\Omega_{c}\rho_{32}^*
		-\Gamma_2\rho_{22},\\
\label{eq:OBE_rho33}
	\frac{\partial}{\partial t}\rho_{33} &=&
		-\frac{i}{2}\Omega_{p}^*\rho_{31} +\frac{i}{2}\Omega_{p}\rho_{31}^*
		 -\frac{i}{2}\Omega_{c}\rho_{32} 
		\nonumber \\
		&+& \frac{i}{2}\Omega_{c}\rho_{32}^* -\Gamma\rho_{33}, \\
\label{eq:OBE_population}
	\frac{\partial}{\partial t}\rho_{11} &=&
		\frac{i}{2}\Omega_p^*\rho_{31} -\frac{i}{2}\Omega_p\rho_{31}^* 
		+\Gamma\rho_{33} \\
\label{eq:MSE}
	\frac{1}{c}\frac{\partial}{\partial t}\Omega_p
		&+& \frac{\partial}{\partial z}\Omega _p = i\frac{\alpha\Gamma}{2L}\rho_{31},
\label{eq:OBE_last}
\end{eqnarray}
where $\rho_{ij}$ ($i, j = 1,2,3$) represents a matrix element of the density matrix operator, $\Omega_{p}$ and $\Omega_c$ are the Rabi frequencies of the probe and coupling fields, $\delta$ is the two-photon detuning, $\gamma_0$ is the intrinsic decoherence rate in the experimental system, $\Delta_p$ and $\Delta_c$ are the one-photon detunings of the probe and coupling transitions, and $\alpha$ and $L$ are the optical depth and length of the medium. 

The values of $\gamma_{\rm DDI}$ and $\delta_{\rm DDI}$ were obtained by the analytic formulas of the attenuation coefficient ($\Delta \beta$) and the phase shift ($\Delta \phi$), respectively, induced by the DDI in the steady-state condition. The formulas were derived in Ref.~\cite{OurOE2020}, utilizing a mean-field model based on the nearest neighbor distribution, and they were experimentally verified in Ref.~\cite{OurCommunPhys2021}. At the condition of $\delta = 0$ and $\Delta_c \ll \Gamma$, the formulas of $\Delta \beta$ and $\Delta \phi$ are given by:
\begin{equation}
\label{eq:Deltabeta}
	\Delta\beta \left( \equiv \frac{2 \alpha \Gamma \gamma_{\rm DDI}}{\Omega_c^2} \right)
	 = \frac{2 \pi^2 \alpha \sqrt{|C_6| \Gamma}}{3 \Omega_c}  n_{R},
\end{equation}

\begin{equation}
\label{eq:Deltaphi}
	\Delta\phi \left(\equiv \frac{\alpha \Gamma \delta_{\rm DDI}}{\Omega_c^2} \right)
		= \frac{\pi^2 \alpha \sqrt{|C_6| \Gamma}}{3 \Omega_c} n_{R},
\end{equation}
where $n_R$ is the density of the Rydberg-state atoms. Using the relation $n_R = n_{a}\rho_{22}$ between $n_R$ and the density of all atoms, $n_a$, we introduced the coefficient $A$, which is written as:
\begin{equation}
\label{definitionA}
	A = \frac{\pi^2 \Omega_c \sqrt{|C_6|}}{3 \sqrt{\Gamma}} n_a.
\end{equation}
The coefficient $A$ represents the DDI-induced decoherence rate or frequency shift per unit $\rho_{22}$. From Eqs.~(\ref{eq:Deltabeta}) and (\ref{definitionA}), we obtained: 
\begin{equation}
	\gamma_{\rm DDI}= A \rho_{22}.
\end{equation}
Similarly, from Eqs.~(\ref{eq:Deltaphi}) and (\ref{definitionA}), we also obtained:
\begin{equation}
	\delta_{\rm DDI}= A \rho_{22}.
\end{equation}
Thus, $\gamma_{\rm DDI}$ and $\delta_{\rm DDI}$ were replaced by $A \rho_{22}$ in the OBEs. According to the values of $\Omega_c$, $C_6$, and $n_a$ used in the experiment, it was determined that $A =$ 0.76$\Gamma$. We set and fixed $A$ to 0.76$\Gamma$ in all the theoretical predictions of this work and the values of $\gamma_{\rm DDI}$ and $\delta_{\rm DDI}$ are linearly proportional to the population of the Rydberg state, $\rho_{22}(z,t)$.

Furthermore, the effect of the DDI, i.e., $\gamma_{\rm DDI}$ and $\delta_{\rm DDI}$, was ignored to obtain Eq.~(\ref{eq:OBE_rho32}) due to the assumption of $A\rho_{22}, \Gamma_2 \ll \Gamma$, which was the case in the experiment. Throughout this study, the two-photon resonance condition of $\delta$ $(\equiv \Delta_p + \Delta_c) = 0$ was always maintained. Please note that the term of $i(\delta_{\rm DDI} + \delta) \rho_{21}$ in Eq.~(\ref{eq:OBE_rho21}), i.e., $i(A\rho_{22} + \delta) \rho_{21}$, played a negligible role in the output probe transmission, since  $\delta = 0$ and $4(A\rho_{22})^2\Gamma^2 \ll \Omega_c^4$ in all the cases of this work. To obtain Eqs.~(\ref{eq:OBE_rho33}) and (\ref{eq:OBE_population}), we also considered that the population in $|2\rangle$ rarely decayed to $|1\rangle$ and $|3\rangle$. When we made the population in $|2\rangle$ all decay to $|1\rangle$ by adding the term of $+\Gamma_2\rho_{22}$ to the right-hand-side of Eq.~(\ref{eq:OBE_population}), the calculation result changed very little because $\Omega_p^2 \ll \Omega_c^2$ in this work.

\subsection{Distorted output pulses of slow light}
\label{subsec:Gaussian1}

Before the measurement, the values of the one-photon detuning and two-photon detuning were set to zero and experimentally verified. We determined the experimental parameters in the order of $\Omega_c$ $\rightarrow$ $\gamma_0$ $\rightarrow$ OD. The parameters were confirmed again after the measurement in the reverse order of that before measurement. Determined $\Omega_c$, $\gamma_0$, and OD were  1.0$\Gamma$, 9.0$\times10^{-3}\Gamma$ and 70, respectively. Details can be found in Supplemental material and Ref.~\cite{OurPRA2019}.

Following the definition of the half mean distance between Rydberg polaritons, $r_a=(4\pi n_{R}/3)^{-1/3}$, the estimated smallest $r_a$ in this experiment was 5.0~$\mu$m when $\Omega_{p0}$ was $0.2$$\Gamma$~\cite{OurOE2020}. The blockade radius, $r_B$, was 1.9~$\mu$m according to the formula of $r_B=(2 |C_6| \Gamma/\Omega_c^2)^{1/6}$~\cite{Lukin_Nature_2012}. $\Omega_{p0}=0.2\Gamma$ was the largest input probe Rabi frequency and $\Omega_c=1.0\Gamma$ was kept throughout the EIT experiment. Thus, it was considered a weakly-interacting many-body system of Rydberg polaritons, which satisfied the condition of $(r_B/r_a)^{3}\ll 1$.

\FigTwo

Using the above experimental conditions, we measured the slow light data as shown in Fig.~\ref{fig:Gaussian1}(a). An input probe pulse with a Gaussian $e^{-1}$ full width of 11.5~$\mu$s was used in this measurement. This input pulse was far longer than that used in the determining the OD. Considering the values of OD and $\Omega_c$, the delay time $\tau_d$ was expected to be $\sim$$1.8~\mu$s as the result of a short input Gaussian pulse with the $e^{-1}$ full width of 0.66~$\mu$s. However, the peak position of the output pulse with $\Omega_{p0}=0.05$$\Gamma$ of the long Gaussian input probe pulse was nearly the same as the peak position of the input pulse (i.e., the delay time was nearly zero). Furthermore, a stronger input pulse had an output peak position that preceded the input peak position and showed a more distorted shape of the output pulse.

The degree of distortion of the output pulse shape depends on $\Omega_{p0}$, or equivalently the Rydberg atom density. One might guess that the DDI effect could explain the distortion, and the experimental data of Fig.~\ref{fig:Gaussian1}(a) could be predicted by Eqs.~(\ref{eq:OBE_first})-(\ref{eq:OBE_last}) with the introduction of $\gamma_{\rm DDI}$ and $\delta_{\rm DDI}$. To test this hypothesis, we calculated the theoretical predictions as shown in Fig.~\ref{fig:Gaussian1}(b) by solving Eqs.~(\ref{eq:OBE_first})-(\ref{eq:OBE_last}), which included the effect of DDI. We used the experimentally-determined values of $\Omega_{p0}$, $\alpha$, $\Omega_c$, and $\gamma_0$ in the calculation. Owing to the DDI effect, i.e., the term of $\gamma_{\rm DDI}$ or $A\rho_{22}$ in Eq.~(\ref{eq:OBE_rho21}), we did observe a lower transmission with a higher value of $\Omega_{p0}$ in the theoretical prediction. In each case of $\Omega_{p0}$, the theoretical prediction gave a very similar delay time ($\sim$$1.8~\mu$s), which resulted in a higher peak transmission than the experimental value. The theoretical predictions did not agree with the experimental data.

\subsection{Phenomenon of the DDI-induced attenuation increasing with time}

\FigThree

To further study the phenomenon, we applied the long square probe pulse with the input Rabi frequency $\Omega_{p0}$ of 0.08$\Gamma$, 0.10$\Gamma$, 0.15$\Gamma$, or 0.2$\Gamma$ under the constant presence of the coupling field, as shown in Fig~\ref{fig:SquarePulse1}. The experimental parameters of $\alpha$ (OD) = 70, $\Omega_c$ = $1.0$$\Gamma$, $\Delta$ = 0, and $\delta$ = 0 were kept the same as in the measurements with the Gaussian input, but $\gamma_0$ was changed to $1.1 \times 10^{-2}$$\Gamma$ in the measurement. 

Regarding the behavior of the output probe field in Fig.~\ref{fig:SquarePulse1}, we observed that the transmission decreased (or the attenuation increases) with time, after it reached the peak value in each of the measurements. The peak value of the output probe transmission decreased with $\Omega_{p0}$. This was expected from the DDI effect, i.e., the term of $\gamma_{\rm DDI}$ or $A\rho_{22}$ in Eq.~(\ref{eq:OBE_rho21}), where $\rho_{22} \approx \Omega_p^2/\Omega_c^2$. However, neither the theoretical model nor Eqs.~(\ref{eq:OBE_first})-(\ref{eq:OBE_last}) described in Sec.~\ref{subsec:theorywoaccumulation} were not able to explain the phenomenon of the transmission decay or the increasing attenuation over time. We fitted the decayed part of the experimental data with an exponential-decay function, and determined the decay time constant and the steady-state attenuation coefficient. Blue lines shown in Fig.~\ref{fig:SquarePulse1} are the best fits. 

\FigFour

Based on the best fits in Fig.~\ref{fig:SquarePulse1}, the decay time constant, $\tau_{sq}$, and steady-state attenuation coefficient, $\beta_{ss}$, as functions of $\Omega_{p0}^2$ are shown in Figs.~\ref{fig:DecayConstant}(a) and \ref{fig:DecayConstant}(b), respectively. The value of $\beta_{ss}$ was given by the absolute value of the logarithm of the transmission at the steady state. Figures~\ref{fig:DecayConstant}(a) and \ref{fig:DecayConstant}(b) clearly show that a larger value of $\Omega_{p0}^2$ resulted in a faster decay and a larger steady-state attenuation. 

A higher Rydberg polariton density leads to a larger DDI-induced attenuation. Thus, the data in Figs.~\ref{fig:SquarePulse1} and \ref{fig:DecayConstant} inferred that the Rydberg polariton density increased with time. One can guess that some of the Rydberg population in $|2\rangle$ was transferred to other states (denoted as $|4\rangle$), and that the atoms in $|4\rangle$ did not interact with the coupling field but were able to have a DDI with the atoms in $|2\rangle$. Due to the existence of the DDI, $|4\rangle$ had to represent some nearby Rydberg states of $|2\rangle$. The population in $|4\rangle$, which was not driven by the coupling field, should have accumulated. As more population accumulated in $|4\rangle$, the DDI-induced attenuation of $|2\rangle$ became large. Thus, the probe output transmission decreased with time, exhibiting the accumulation phenomenon of the DDI effect.

In Fig.~\ref{fig:EIT_transition}(a), we introduce $|4\rangle$ to represent all possible nearby Rydberg states of $|2\rangle$. The spontaneous or collisional decay process could cause the transfer of the population from $|2\rangle$ to $|4\rangle$, because no additional field was applied in the experiment. Since $|2\rangle$ was driven by the coupling field, it was called the bright Rydberg state. On the other hand, $|4\rangle$ did not interact with any applied field and was called the dark Rydberg state.

\subsection{Evidence of dark Rydberg states}
\label{subsec:Gamma4}

To verify the hypothesis based on the dark Rydberg state $|4\rangle$, we measured the existing time of the DDI effect, after nearly all the atoms in the bright Rydberg state $|2\rangle$ had been de-excited. The basic idea of this measurement was as follows. 
Since the atoms in $|4\rangle$ did not interact with any applied field, they should decay by themselves and have a decay rate of $\Gamma_4$. Furthermore, $|4\rangle$ represented a number of Rydberg states, so $\Gamma_4^{-1}$ must be close to the Rydberg-state lifetime. Consequently, after the atoms in $|2\rangle$ had been removed, the DDI effect should still exist in the system, and gradually decay with the time constant of $\Gamma_4^{-1}$. Then, we explored the DDI-induced attenuation as a function of time in the system by employing a very weak Gaussian probe pulse and the coupling field. The probe pulse was weak enough to cause a negligible DDI effect by itself. 

The sequence of the input probe field is depicted in the inset of Fig.~\ref{fig:PollutantDecayTime}. 
A 70-$\mu$s square probe pulse of $\Omega_{p0}$ = 0.1$\Gamma$ was first employed. The pulse duration was sufficiently long compared with the lifetime of $|2\rangle$, which was approximately $20~\mu$s at room temperature. The OD was 70 and the coupling field of $\Omega_c =$ 1.0$\Gamma$ was continuously kept on during the measurement. After the square probe pulse was switched off, the coupling field quickly de-excited the remaining population in $|2\rangle$. The de-excitation time was estimated to be about 27~ns. We waited for a certain time $\Delta t$ after the square pulse was turned off, and then applied a weak Gaussian probe pulse to measure the DDI-induced attenuation, $\beta_G$, as a function of $\Delta t$. The value of $\beta_G$ was equal to the absolute value of the logarithm of the peak transmission of the output Gaussian pulse. The input Gaussian pulse had the peak $\Omega_{p0}$ of 0.05$\Gamma$ and the $e^{-1}$ full width of 10~$\mu$s. Compared with the 70-$\mu$s square probe pulse, the weaker and much shorter Gaussian probe pulse induced a negligible DDI effect. 

\FigFive

We observed the exponential-decay behavior of $\beta_G$ versus $\Delta t$ as shown in Fig.~\ref{fig:PollutantDecayTime} in which the circles are the experimental data and the black line is the best fit of an exponential-decay function. The decay time constant of the best fit was 31 $\mu$s, which is about the lifetime of a Rydberg state with a principal quantum number between 32 and 38. The atoms in $|2\rangle$ were quickly de-excited by the coupling field, after the square probe pulse was switched off. If the atoms in the dark Rydberg state $|4\rangle$ could not exist in the system, the value of $\beta_G$ would quickly drop to its steady-state value due to the absence of the atoms in the bright Rydberg state $|2\rangle$. The slow decay of $\beta_G$ demonstrated the existence of the atoms in $|4\rangle$, which did not interact with the coupling field, but gave the DDI effect and decayed slowly.

\section{Theory of the accumulation effect and experimental verification}
\label{sec:modifiedtheory}

\subsection{Theoretical model with the accumulation effect}
\label{thwithaccumulation}


In the previous section, we discussed the necessity of the consideration of dark Rydberg states in the system. To simulate the accumulative DDI effect observed in the experiment, we improved the theoretical model described in Subsec.~\ref{subsec:theorywoaccumulation} by including the dark Rydberg state $|4\rangle$ and the decay process from $|2\rangle$ to $|4\rangle$. The DDI effect between one atom in $|2\rangle$ and one atom in $|4\rangle$ was also taken into account. Thus, the OBEs can be given by:
\begin{eqnarray}
\label{eq:mOBE_first}
\label{eq:mOBE_rho21}
	\frac{\partial}{\partial t}\rho_{21} &=& 
		\frac{i}{2}\Omega_{c}\rho_{31} +i (A\rho_{22} + B'\rho_{44} + \delta) \rho_{21}
		\nonumber \\
		&-&\left( A\rho_{22} +B\rho_{44} +\gamma_0 +\frac{\Gamma_{24}}{2} \right) 
		\rho_{21}, \\
\label{eq:mOBE_rho31}		
	\frac{\partial}{\partial t}\rho_{31} &=&
		\frac{i}{2}\Omega_{p} (\rho_{11}-\rho_{33})
		+\frac{i}{2}\Omega_{c}\rho_{21} +i\Delta_{p}\rho_{31} \nonumber \\
		&-&\frac{\Gamma}{2}\rho_{31}, \\
\label{eq:mOBE_rho32}
	\frac{\partial}{\partial t}\rho_{32} &=&
		\frac{i}{2}\Omega_{p}\rho_{21}^* +\frac{i}{2}\Omega_{c}(\rho_{22}-\rho_{33})
		\nonumber \\
		&-& i\Delta_c \rho_{32}-\frac{\Gamma}{2} \rho_{32}, \\
\label{eq:mOBE_rho22}
	\frac{\partial}{\partial t}\rho_{22} &=&
		\frac{i}{2}\Omega_{c}\rho_{32}-\frac{i}{2}\Omega_{c}\rho_{32}^*
		-\Gamma_{24}\rho_{22}, \\
\label{eq:mOBE_rho33}
	\frac{\partial}{\partial t}\rho_{33} &=&
		-\frac{i}{2}\Omega_{p}^*\rho_{31} +\frac{i}{2}\Omega_{p}\rho_{31}^*
		 -\frac{i}{2}\Omega_{c}\rho_{32} 
		\nonumber \\
		&+& \frac{i}{2}\Omega_{c}\rho_{32}^* -\Gamma\rho_{33}, \\
\label{eq:mOBE_rho44}
	\frac{\partial}{\partial t}\rho_{44} &=& \Gamma_{24}\rho_{22} -\Gamma_4\rho_{44},\\
\label{eq:mOBE_population}
	\frac{\partial}{\partial t}\rho_{11} &=&
		\frac{i}{2}\Omega_p^*\rho_{31} -\frac{i}{2}\Omega_p\rho_{31}^* 
		+\Gamma\rho_{33},
\end{eqnarray}
where the coefficient $A$ represents the DDI-induced decoherence rate or frequency shift per unit $\rho_{22}$ with a value given by Eq.~(\ref{definitionA}), the coefficient $B$ (or $B'$) is similar to the coefficient $A$ and represents the DDI-induced decoherence rate (or frequency shift) per unit $\rho_{44}$, $\Gamma_{24}$ is the decay rate from $|2\rangle$ to $|4\rangle$, and $\Gamma_4$ is the decay rate of $|4\rangle$. 

The terms $A\rho_{22} + B'\rho_{44}$ and $A\rho_{22} + B\rho_{44}$ in Eq.~(\ref{eq:mOBE_rho21}) represent $\delta_{\rm DDI}$ and $\gamma_{\rm DDI}$, respectively. The values of $\gamma_{\rm DDI}$ and $\delta_{\rm DDI}$ varied with space and time due to the populations of $\rho_{22}(z,t)$ and $\rho_{44}(z,t)$ during the propagation of the probe light. The decay rate $\Gamma_{24}$ was parametrized as:
\begin{equation}
\label{eq:Gamma24}
	\Gamma_{24} \equiv C\rho_{22} + D,
\label{eq:mOBE_last}
\end{equation}
where the coefficient $C$ is the two-body decay rate per unit $\rho_{22}$, and the coefficient $D$ is the one-body decay rate. The atomic density, $n_a$, of the system is a part of the coefficient $C$. Further discussion regarding the transfer mechanism can be found in the Discussion section. 

In Eq.~(\ref{eq:mOBE_rho32}), the DDI effect, i.e., $\gamma_{\rm DDI}$ and $\delta_{\rm DDI}$, and the decay rates, i.e., $\Gamma_{24}$, and $\Gamma_2$, were ignored due to the experimental condition of $A\rho_{22}, B\rho_{44}, \Gamma_{24}, \Gamma_2 \ll \Gamma$. Please note that the frequency shift induced by the population in each of the dark Rydberg states might be positive or negative, and thus the net frequency shift resulted in $|B'| \leq B$. Furthermore, the term of $i(A\rho_{22} + B'\rho_{44} + \delta) \rho_{21}$ in Eq.~(\ref{eq:mOBE_rho21}) played a negligible role in the output probe transmission, since $\delta = 0$ and $4|A\rho_{22} + B\rho_{44}|^2\Gamma^2 \ll \Omega_c^4$ in all the cases of this work.

The value of $A$ given by Eq.~(\ref{definitionA}) was derived from Ref.~\cite{OurOE2020} and experimentally verified in Ref.~\cite{OurCommunPhys2021}. We first measured the coefficients $C$ and $D$ as described in Subsec.~\ref{subsec:Gamma24}. Using the known values of $A$, $C$, and $D$, we then determined $B$ using the experimental data of Fig.~\ref{fig:SquarePulse1} as illustrated in Subsec.~\ref{subsec:comparison}. Finally, we compared the experimental data of the slow light shown in Fig.~\ref{fig:Gaussian1} with the theoretical predictions calculated using the experimentally determined values of $A$, $B$, $C$, and $D$ in Subsec.~\ref{subsec:comparison}. The comparison was used to test the validity of the theoretical model introduced here.

\subsection{Determination of the decay rate from bright to dark Rydberg states}
\label{subsec:Gamma24}

We designed an experiment to determine the coefficients $C$ and $D$ in Eq.~(\ref{eq:Gamma24}). The decay rate, $\Gamma_{24}$, of the atoms in $|2\rangle$ was measured against the atomic density of $|2\rangle$, $n_R$. The experiment details can be found in Supplemental material. The procedure of the measurement of $\Gamma_{24}$ was as follows. We first prepared a given number of atoms in the ground state $|1\rangle$ and excited them to $|2\rangle$ with a two-photon transition (TPT) pulse. The TPT had a large one-photon detuning $\Delta$. After the TPT pulse, no light field was applied, and the atoms in $|2\rangle$ decayed. Then, we waited for a certain time $\Delta t$, and also depleted the residual atoms in $|1\rangle$. At the end of $\Delta t$, the remaining atoms in $|2\rangle$ were brought back to $|1\rangle$ with another TPT pulse. Finally, after the second TPT pulse we determined the number of the returned atoms in $|1\rangle$ or, equivalently, measured the absorption coefficient, $\Delta \beta$, of the weak probe field. The result of $\Delta \beta$ was proportional to the remaining atoms in $|2\rangle$ after the given $\Delta t$. By varying $\Delta t$ and measuring $\Delta \beta$, we obtained the data points of $\Delta \beta$ as a function of $\Delta t$, and fitted the data with an exponential function. The best fit gave the value of $\Gamma_{24}$.

\FigSix

The circles in Fig.~\ref{fig:Gamma24} are the experimental data of $\Gamma_{24}$ as a function of $n_R$. Each circle represents the result of a series of measurements similar to those shown in Supplemental Material. We fitted the circles in Fig.~\ref{fig:Gamma24} with the fitting function of a straight line. The interception of the best fit determined the coefficient $D$ defined in Eq.~(\ref{eq:Gamma24}), which was 2.0$\times$$10^{-3}$$\Gamma$. The determined value of $D$ was comparable to the spontaneous decay rate of $32D_{5/2}$ at room temperature, which is $1.3$$\times$$10^{-3}$$\Gamma$~\cite{Rb87Data, RydbergStateLifeTime}. Furthermore, the slope, $k$, of the best fit determined the coefficient $C$ defined in Eq.~(\ref{eq:Gamma24}) in the following way: Since $k \, n_R = C \rho_{22}$ and $n_R = \rho_{22} n_a = \rho_{22} \alpha/(\sigma_0 L)$, the coefficient $C$ is given by:
\begin{equation}
	C = k \alpha/(\sigma_0 L),
\end{equation}
where $\alpha$ is the OD used in the EIT experiment, $\sigma_0$ is the absorption cross section of the resonant probe transition from $|5S_{1/2}, F=2, m_F=2\rangle$ to $|5P_{3/2}, F=3, m_F=3\rangle$. During the measurements of the data shown in Figs.~2(a) and 3(a)-3(d), the optical depth ($\alpha$) was 70, we used the above equation to estimate that the value of $C$ was $9.1$$\times$$~10^{-2}$$\Gamma$ at $\alpha =$ 70. Possible processes of the population transfer from the bright to dark Rydberg states are discussed in Sec.~\ref{sec:discussion}.

\subsection{Comparison between theoretical predictions and experimental results}
\label{subsec:comparison}

\FigSeven

After determining the values of $C$ and $D$ in the decay rate $\Gamma_{24}$ used in Eqs.~(\ref{eq:mOBE_rho21}), (\ref{eq:mOBE_rho22}), and (\ref{eq:mOBE_rho44}), we next utilized the experimental data of Fig.~\ref{fig:SquarePulse1} to determine the value of coefficient $B$ used in Eq.~(\ref{eq:mOBE_rho21}). Since the measured value of $\Gamma_{24}$ was the total decay rate of $|2\rangle$ but not merely the decay rate from $|2\rangle$ to $|4\rangle$, the determined value of $B$ might account for the discrepancy. Figure~\ref{fig:SquarePulse2} shows the comparison between the experimental data of the long square probe pulses shown in Figs.~\ref{fig:SquarePulse1}(a)-\ref{fig:SquarePulse1}(d) and the theoretical predictions calculated with the modified model described in Sec.~\ref{thwithaccumulation}. In the calculation, we used the experimentally determined values of the optical depth, the coupling Rabi frequency, and intrinsic decoherence rate $\gamma_0$ in the system \cite{OurPRA2019}. We set $A$ (the DDI-induced decoherence rate per unit $\rho_{22}$) to the value given by Eq.~(\ref{definitionA}), and used the values of $C$ and $D$ derived from the result of Fig.~\ref{fig:Gamma24}. The only adjustable parameter in the calculation of the predictions was $B$ (the DDI-induced decoherence rate per unit $\rho_{44}$). Note that we set $B'$ = $B$ in the term of $i(A\rho_{22} + B'\rho_{44} + \delta) \rho_{21}$ in Eq.~(\ref{eq:mOBE_rho21}), but this term played a negligible role in the calculation result of the output probe transmission. A good agreement between the theoretical predictions and experimental data was obtained at the coefficient $B  =$ 7.7$~\Gamma$.

A single optimum value of $B$ was able to explain all the experimental data taken at different input probe Rabi frequencies. This demonstrated the theoretical model presented by Eqs.~(\ref{eq:mOBE_first})-(\ref{eq:mOBE_last}) was qualitatively valid. Furthermore, we compared the experimental data of the Gaussian input probe pulses as shown in Fig.~\ref{fig:Gaussian1} with the theoretical predictions. The values of the parameters used in the calculation of the predictions were the same as those in Fig.~\ref{fig:SquarePulse2} except for the value of $\gamma_0$, which had a day-to-day fluctuation of 1.5$\times$$10^{-3}$$\Gamma$. Figures~\ref{fig:Gaussian2}(a)-\ref{fig:Gaussian2}(c) show that the experimental data were all in good agreement with the theoretical predictions, manifesting that the theoretical model was also quantitatively successful.

\FigEight

\section{Discussion}
\label{sec:discussion}

We discuss some possible mechanisms of the population transfer to the dark Rydberg state in this section. Experimental observations led by the existence of the dark Rydberg state which was transferred from a bright Rydberg state have been reported in several articles \cite{Martin_2004, Porto_2016,Porto_2017, Hond_2020, Gorshkov_2020,PRA_SR_2007,NJP_SR_2021}. The underlying mechanisms of such transfers can be classified into four as follows: (i) transitions driven by a microwave field \cite{Martin_2004}, (ii) the spontaneous decay \cite{Porto_2016,Porto_2017,Hond_2020}, (iii) the DDI-induced antiblockade excitation and state-change collision assisted by radiation trapping \cite{Gorshkov_2020,radiationtrapping}, and (iv) the superradiance of the transition with a long wavelength induced by the black-body radiation (BBR)~\cite{PRA_SR_2007,NJP_SR_2021}. Please see Supplemental material for further discussion on these articles.

However, the mechanisms (i), (ii), and (iii) cannot explain our observations. Corresponding reasons are following. (a) No additional microwave field was applied for population transfer to another Rydberg state. (b) The spontaneous decay was a one-body process, and its rate should not depend on the Rydberg-atom density. However, the observed decay rate of the population in the bright Rydberg state depended on the Rydberg-atom density, and it is much higher than the spontaneous decay at the room temperature. (c) According to the experimental condition, we made an estimation, illustrated in the next paragraph, to show that the state-changing Rydberg collision rate~\cite{state_changing_collision_PRA2015,PRL_stateMixing2004,PRL_stateMixing2008,PRR_stateMixing2020} is too low to be responsible for the population transfer from the bright to dark Rydberg states. Furthermore, the antiblockade excitation~\cite{antiblockade_PRL2007,antiblockade_PRL2010} is difficult to occur in the weak-interaction regime like our experiment due to large energy defects between  and nearby states.

We can rule out the mechanisms (i) and (ii) for the corresponding quite obvious reasons (a) and (b). However, before rule out the third case, we estimated the DDI induced state-exchange collision or the Rydberg state-changing collision rate compared to Ref.~\cite{Gorshkov_2020}. The decay rate due to state-changing collisions is proportional to $n_R \times$(principal quantum number)$^{12}$~\cite{state_changing_collision_PRA2015}. Comparing the experimental conditions of the present study ($n_R$$\sim$$10^{9}~$cm$^{-3}$, principal quantum number = 32) with Ref.~\cite{Gorshkov_2020} 
($n_R$$\sim$$10^{9}~$cm$^{-3}$, principal quantum number = 111), the state-changing collision rate in our experiment should be on the order of 0.1~Hz based on the rate presented in Ref.~\cite{Gorshkov_2020}. The estimated value is far smaller than the determined $\Gamma_{24}$ shown in Fig.~\ref{fig:Gamma24}. Therefore, the state-changing collision is not responsible for the observed two-body decay.

As a possible mechanism of the population transfer from the bright to dark Rydberg states, we considered the superradiance of transitions induced by the BBR. According to the transition or decay rate from $|32D_{5/2}\rangle$ to the nearby states induced by the BBR at $T=300$~K and the corresponding wavelength, we found the four most probable dark Rydberg states. Their information is listed in Table~\ref{t:superradiance}. To explain the measured two-body decay rate by the superradiance of the transitions induced by the BBR at the room temperature, we made the estimation the decay rate from $|32D_{5/2}\rangle$ to $|33P_{3/2}\rangle$ as an example. In Fig.~\ref{fig:Gamma24}, the decay rate at $n_R=0.003$$~\mu$m$^{-3}$ was about five times faster than the spontaneous decay rate of $|32D_{5/2}\rangle$ at the room temperature. To estimate the number of particles participating in the superradiance, i.e., the enhancement factor of the decay rate, we considered an effective range for the cooperative interaction to be $\sim$1/100 of the transition wavelength ($\lambda_0$)~\cite{PRA_SR_2007,NJP_SR_2021}. The number of Rydberg atoms inside the effective volume of $(\lambda_0/100)^3$ was 260 according to the value of $\lambda_0$ in Table~\ref{t:superradiance} and $n_R=0.003~\mu$m$^{-3}$. Hence, the superradiance of the BBR-induced transitions can likely be the mechanism that transfers the population dark Rydberg states faster.

\begin{table}
\caption{The four dark Rydberg states, which the population in $|32D_{5/2}\rangle$ predominately decays to, due to the BBR at the room temperature. $\Gamma_{BBR}$ is the transition rate induced by the BBR at $T=300$~K. $\lambda_0$ and $\Delta E/h$ are the transition wavelength and frequency. $\tau_{ds}$ is lifetime of each state at $T=300$~K. All the values were calculated based on Ref.~\cite{ARC}.}
\begin{tabular}{c c c c c}
 \hline\hline
 State & $\Gamma_{BBR}$ & $\lambda_0$  & $\Delta E/h$ & $\tau_{ds}$ \\ 
 $|nL_J\rangle$    &        (KHz)            &      (mm)          &     (GHz)  & ($\mu$s) \\
 \hline 
 $|30F_{7/2}\rangle$ & 22.5 & 1.9 & 158.4 & 15.0\\ 

 $|31F_{7/2}\rangle$ & 15.1 & 4.0 & 74.0 & 16.4\\ 
 
 $|34P_{3/2}\rangle$ & 14.8 & 1.9 & 155.4 & 34.7\\  
 
 $|33P_{3/2}\rangle$ & 9.1 & 4.4 & 68.63 & 32.2\\ 
 \hline\hline

\end{tabular}
\label{t:superradiance}
\end{table}

The values of $A$ and $B$ were $0.76~\Gamma$ and $7.7~\Gamma$, respectively, indicating that the DDI strength between a bright and a dark Rydberg atoms is much larger than that between two bright Rydberg atoms. To explain this, we considered the DDI between a bright and a dark Rydberg atoms is a dipole-dipole interaction of the collision process of $|32D_{5/2}\rangle+|nL_J\rangle\rightarrow|nL_J\rangle+|32D_{5/2}\rangle$, where $|nL_J\rangle$ is one of the dark Rydberg state in Table~\ref{t:superradiance}. The collision process is resonant, because the two Rydberg atoms just exchange their quantum states. To estimate the strength of each resonant DDI process between a bright and a dark Rydberg atoms in Table~\ref{t:superradiance}, we followed steps similar to the estimation of $A$ described in Subsec.~\ref{subsec:theorywoaccumulation} and Ref.~\cite{OurOE2020}. The estimated value of $B$ is denoted as $\widetilde{B}$. The relation between $\widetilde{B}$ and the dipolar coupling coefficient $C_3$ is given by
\begin{equation}
\label{eq:estimateB}
	\widetilde{B} = \frac{2 \pi^2 \overline{C_3}}{3} n_a,
\end{equation}
where $\overline{C_3}$ is the average value of $C_3$ by considering the branch ratio as a weight value. The branch ratio is the ratio between the decay or excitation rates of the $\Delta m_J = \pm 1, 0$ transitions from $|32D_{5/2}\rangle$ to $|nL_J\rangle$. The estimated value of $\widetilde{B}$ with $\overline{C_3}$ of each possible dark Rydberg state was listed in Table~\ref{t:C3}.
 
Although the simple derivation of using Eq.~(\ref{eq:estimateB}) may cause overestimation of $\widetilde{B}$, a resonant dipole-dipole interaction between the bright and dark Rydberg states can provide an intuitive understanding of the behavior of the system. The resonant DDI strength arises from $C_3$, and $\overline{C_3}$ is the average value of $C_3$ among the Zeeman states of the same $|nL_J\rangle$. The strength of the non-resonant DDI between two bright Rydberg atoms arises from $C_6$. According to Eqs.~(\ref{definitionA}) and (\ref{eq:estimateB}), we can obtain the ratio of 
$A/\widetilde{B}=(\Omega_c \sqrt{|C_6|/\Gamma}/(2\overline{C_3})$. Since $\overline{C_3}$ in Table~\ref{t:C3} is much larger than $(\Omega_c \sqrt{|C_6|/\Gamma}/2)=2\pi\times14$~MHz$\cdot\mu$m$^3$ in the experiment, $\widetilde{B}$ or the estimated value of $B$ is much larger than $A$. Therefore, the experimentally-determined value of $B$ indicated the interaction between a bright and a dark Rydberg atoms is likely a resonant DDI. Lifetime of each of the four possible dark Rydberg states is also listed in Table~\ref{t:superradiance}.
 The lifetimes of $|30F_{7/2}\rangle$ and $|31F_{7/2}\rangle$ are significantly shorter than those of $|33P_{3/2}\rangle$ and $|34P_{3/2}\rangle$. The population accumulated in the former is far less than that in the latter. Therefore, we think the dark Rydberg states of $|33P_{3/2}\rangle$ and $|34P_{3/2}\rangle$ contribute the coefficient $B$ much more than $|30F_{7/2}\rangle$ and $|31F_{7/2}\rangle$. The decay time constant, i.e., $31~\mu$s, of the best fit in Fig.~\ref{fig:PollutantDecayTime} is close to the lifetimes of $|33P_{3/2}\rangle$ and $|34P_{3/2}\rangle$ in Subsec.~\ref{subsec:Gamma4}.

\begin{table}
\caption{Estimation of the strength of the resonant DDI between an atom in state $|32D_{5/2}\rangle$ and another in state $|nL_J\rangle$. $\widetilde{B}$ represents the estimated value of the coefficient $B$ according to Eq.~(\ref{eq:estimateB}), $\overline{C_{3}}$ is the dipolar coupling coefficient, and $\overline{C_{3}}$ is the average value of $C_3$ weighted by the branch ratio, which is the ratio between the decay or excitation rates of the transitions of $\Delta m_{J}=\pm1, 0$ from $|32D_{5/2}\rangle$ to $|nL_J\rangle$. The values of branch ratio and $C_3$ were calculated based on Ref.~\cite{ARC}.}
\begin{tabular}{c c c c c c }
\hline \hline
State &~ ~$m_J$~ &~ Branch ~ & $C_3/(2\pi)$ & $\overline{C_3}/(2\pi)$ & $\widetilde{B}$ \\ 
$|nL_J\rangle$~         &       & ratio   & (MHz$\cdot\mu$m$^3$) &(MHz$\cdot\mu$m$^3$)& ($\Gamma$)  \\
\hline 
                    & 7/2 & 0.58 & 220 &  &  \\ 

$|30F_{7/2}\rangle$ & 5/2 & 0.30 & 63 & 146 & 7.9\\ 

                    & 3/2 & 0.12 & 10.5 &  &  \\ 
\cline{2-4}

                    & 7/2 & 0.57 & 687 &  &  \\ 

$|31F_{7/2}\rangle$ & 5/2 & 0.31 & 196 & 456 & 24.7\\ 

                    & 3/2 & 0.12 & 33 &  &  \\ 
\cline{2-4}
$|34P_{3/2}\rangle$ & 3/2 & 1 & 206 & 206 & 11.2\\ 
$|33P_{3/2}\rangle$ & 3/2 &  1 & 636 & 636 & 34.4\\ 
\hline\hline
\end{tabular} 
\label{t:C3}
\end{table}

\section{Conclusion}
\label{sec:conclusion}

In this work, we systematically studied the transmission of a probe field propagating through a Rydberg EIT system, in which the DDI strength was in the weak interaction regime. We observed the distorted output of a long Gaussian input probe pulse. Such a phenomenon was unable to be predicted by the theoretical model, which only considered the DDI between the bright Rydberg atoms. According to the further measurements, we explained that the distortion was due to the extra attenuation. This phenomenon was caused by a much larger DDI-induced decoherence rate due to the atoms accumulating in the dark Rydberg states. The population in the dark Rydberg states was transferred from the bright Rydberg state in an unexpected high rate. We attribute the high transfer rate to the superradiance of transitions induced by the black-body radiation. Using a theoretical model to take into account the effect that population accumulates in some dark Rydberg states and the dark Rydberg atoms cause an extra DDI-induced decoherence rate, we can successfully explain the experimental data.

In conclusion, this study provided a better understanding for the creation of dark Rydberg atoms and their influence to Rydberg polaritons in the Rydberg-EIT system under a long interaction time~\cite{SLP_Lukin,SPL_2009,SLP_review} and we believe the effect is universal. This work also points out an obstacle in the realization of the BEC with weakly-interacting Rydberg polaritons~\cite{DSP_BEC_Fleischhauer2008,OurCommunPhys2021}.


\section*{Acknowledgements}
We thank the Referees for their comments which stimulate the discussions in Sec.~\ref{sec:discussion}. This work was supported by Grant Nos.~109-2639-M-007-001-ASP and 110-2639-M-007-001-ASP of the Ministry of Science and Technology, Taiwan.


\end{document}


\title{
SUPPLEMENTAL MATERIAL\\ \vspace*{\baselineskip}
Increasing decoherence rate of Rydberg polaritons due to accumulating dark Rydberg atoms
}
\author{Ko-Tang Chen,$^1$
Bongjune Kim,$^{1,}$\footnote{Electronic address: {\tt upfe11@gmail.com}}
Chia-Chen Su,$^1$
Shih-Si Hsiao,$^1$
Shou-Jou Huang,$^2$
Wen-Te Liao,$^{2,3,4}$
and
Ite A. Yu,$^{1,4,}$
}
\email{yu@phys.nthu.edu.tw}

\address{$^1$Department of Physics, National Tsing Hua University, Hsinchu 30013, Taiwan \\
$^2$Department of Physics, National Central University, Taoyuan City 320317, Taiwan \\
$^3$Physics Division, National Center for Theoretical Sciences, Taipei 10617, Taiwan\\ 
$^4$Center for Quantum Technology, Hsinchu 30013, Taiwan
}



\maketitle
\vspace*{-3\baselineskip}

\newcommand{\FigSOne}{
	\begin{figure}[t]
	\center{\includegraphics[width=80mm]{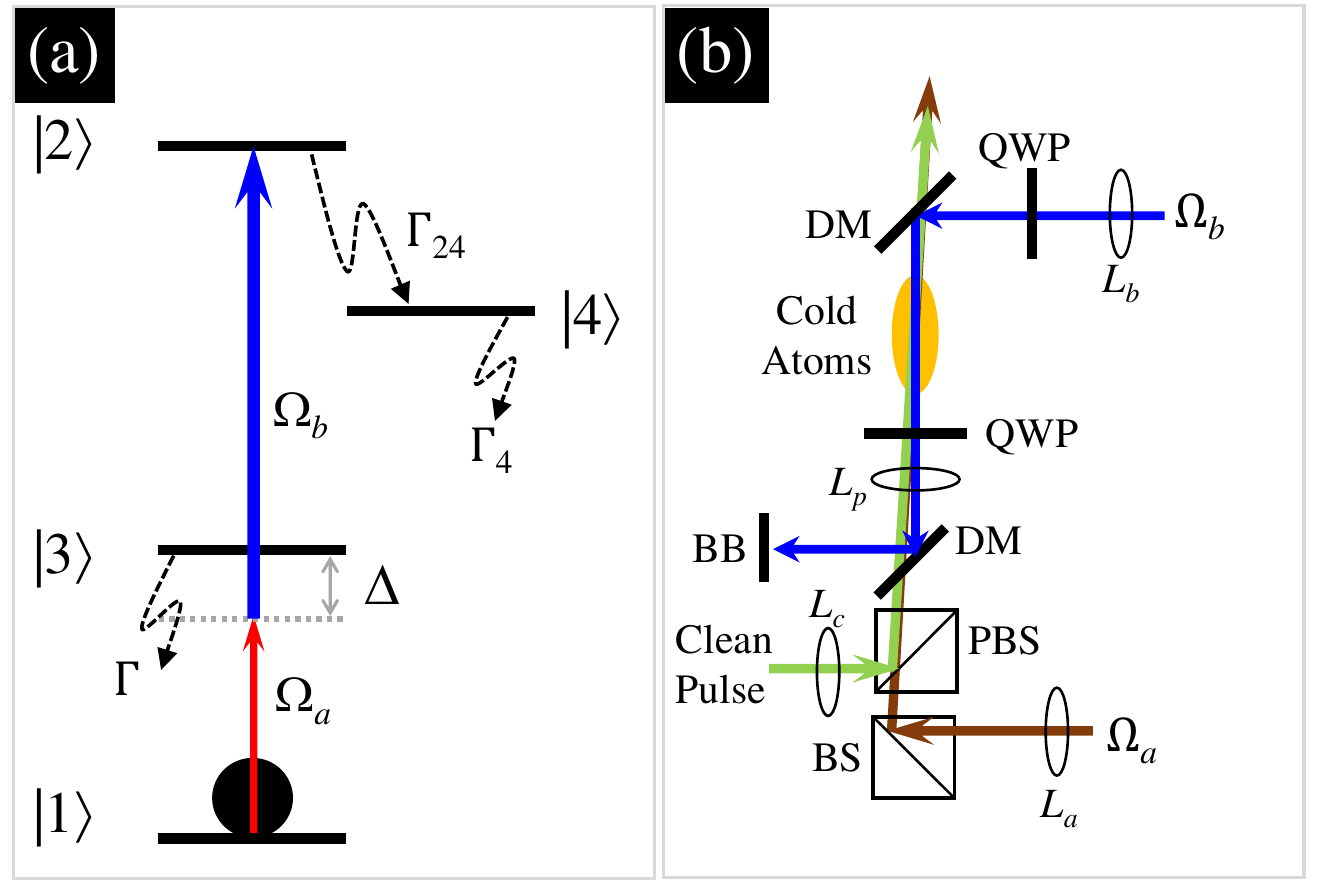}}
	\caption{
(a) Scheme of the two-photon transition, which moved the population from $|1\rangle$ to $|2\rangle$. The energy levels here were the same as those in Fig.~1. Please note that $|1\rangle \rightarrow |3\rangle$ is a cycling transition and, thus, population of any undesired transition to $|3\rangle$ quickly decayed back only to $|1\rangle$. $\Omega_a$ and $\Omega_b$ denote the Rabi frequencies of the two square pulses of the laser fields. We kept the frequencies of $\Omega_a$ and $\Omega_b$ to the two-photon resonance, and applied a one-photon detuning ($\Delta$) of 4$\Gamma$ to $\Omega_b$ or that of $-4$$\Gamma$ to $\Omega_a$. $\Gamma$ and $\Gamma_4$ are the spontaneous decay rates of $|3\rangle$ and $|4\rangle$, respectively. $\Gamma_{24}$ is the decay rate from $|2\rangle$ to $|4\rangle$. (b) Optical paths of $\Omega_a$, $\Omega_b$, and a clean pulse in the experiment. $\Omega_b$ was the same as that of the coupling field, but had a different frequency. The clean pulse was employed to wipe out the population in $|1\rangle$.
}
	\label{fig:Two-Photon_Transition}
	\end{figure}
}
\newcommand{\FigSTwo}{
	\begin{figure}[t]
	\center{\includegraphics[width=60mm]{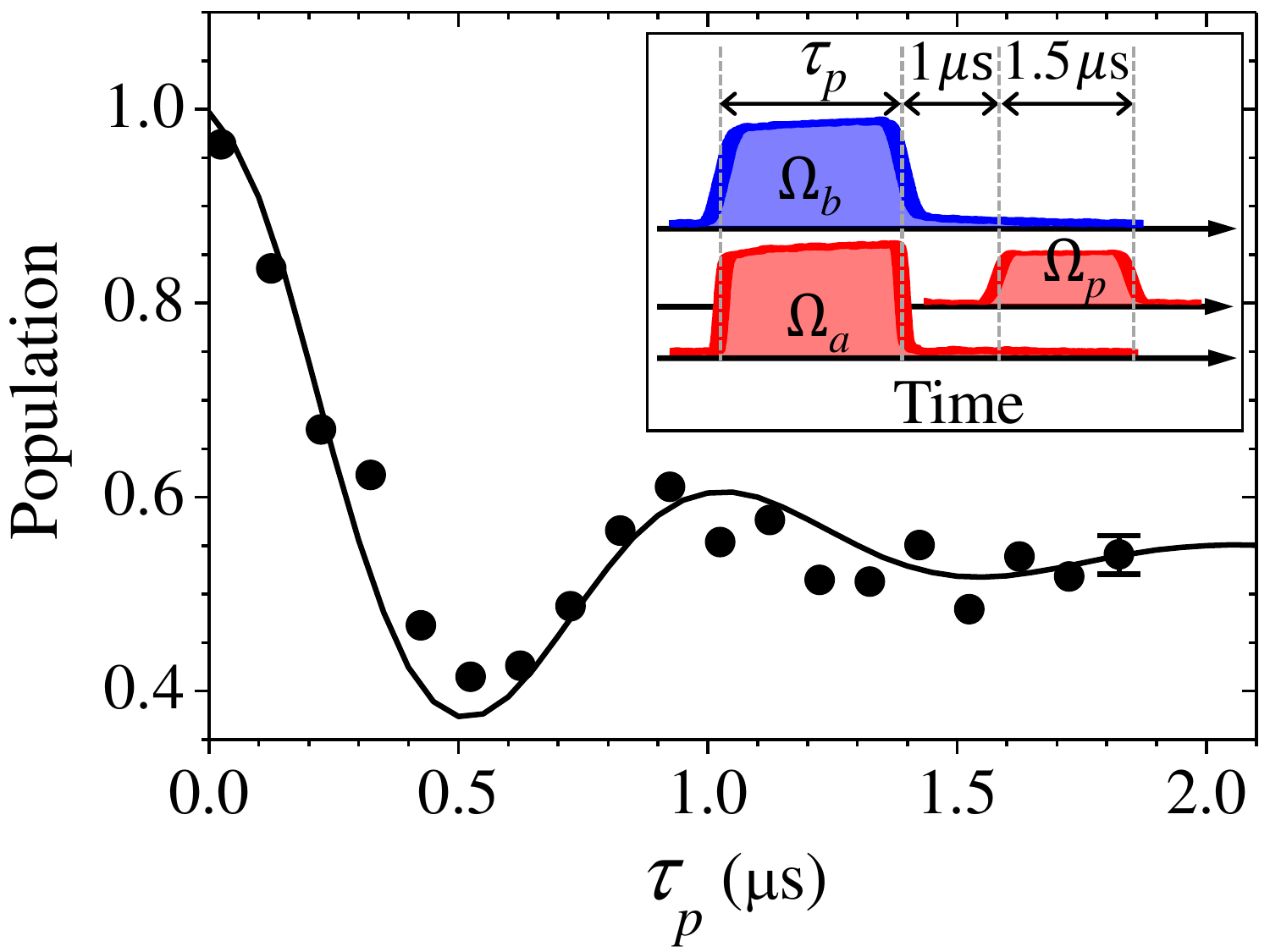}}
	\caption{Experimental demonstration of the two-photon transition (TPT) shown in Fig.~\ref{fig:Two-Photon_Transition}(a). In the main figure, we plot the remaining population in state $|1\rangle$ after the TPT pulse as a function of the TPT pulse width, $\tau_p$. In the inset, we show the timing sequence of the pulses of $\Omega_a$ and $\Omega_b$ (to drive the TPT) and $\Omega_{p'}$ (to measure the population in $|1\rangle$), as well as their pulse shapes. We set $\Delta =$ 4.0$\Gamma$, $\alpha$ (optical depth) = 0.5, $\Omega_a = \Omega_b =$ 1.2$\Gamma$, and $\Omega_{p'} =$ 0.08$\Gamma$ in the measurement. Circles are the experimental data, which have similar error bars. Only the last data point shows the error bar, which represents the typical value. The black line is the theoretical prediction of a simple three-level cascade system under the two-photon resonance. The prediction was calculated with the decoherence rate of 6.5$\times$$10^{-2}$$\Gamma$, and the above values of $\Delta$, $\Omega_a$, and $\Omega_b$. 
}
	\label{fig:TPE}
	\end{figure}
}
\newcommand{\FigSThree}{
	\begin{figure}[t]
	\center{\includegraphics[width=61mm]{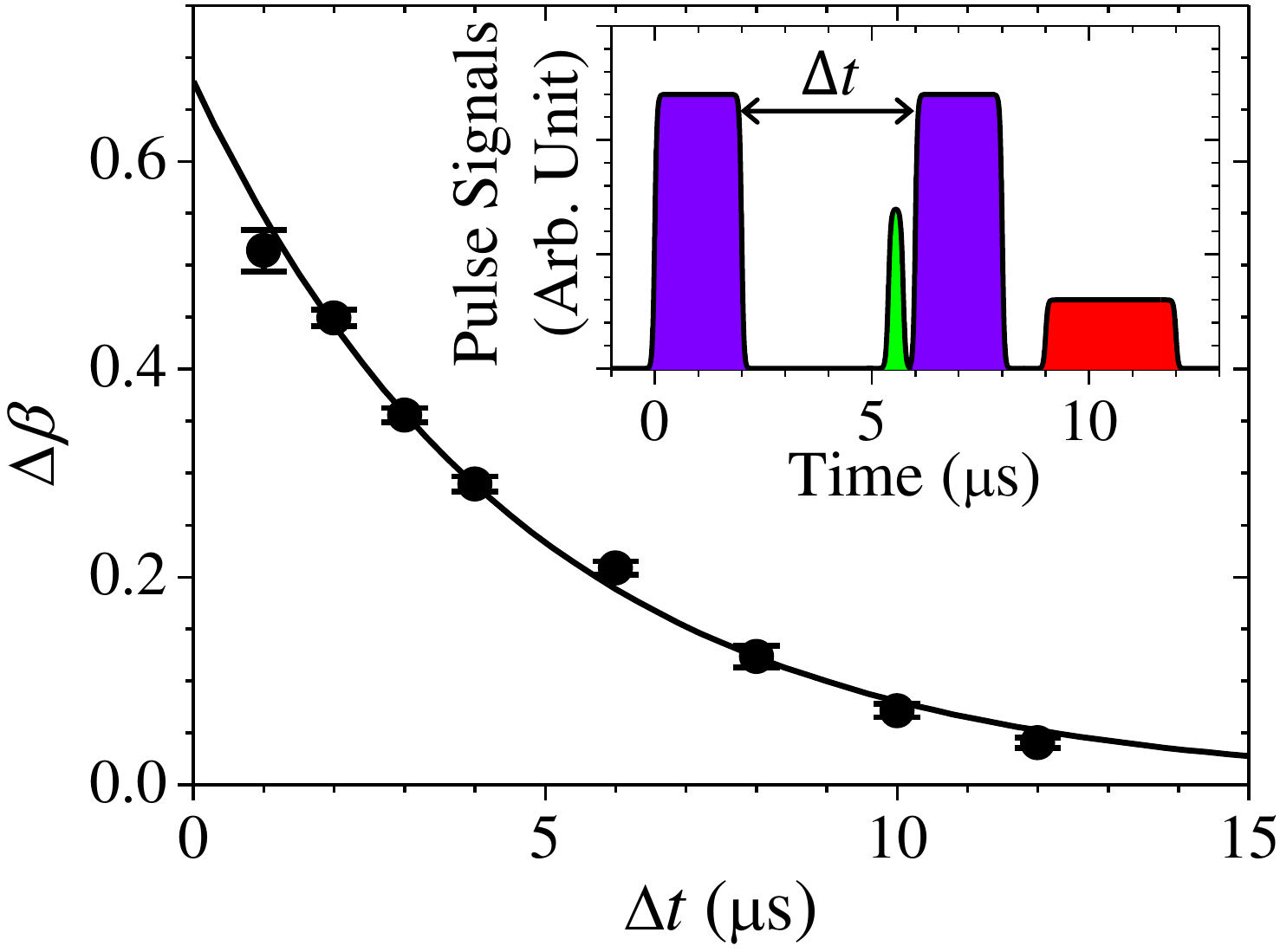}}
	\caption{Determination of the decay rate of the bright Rydberg state $|2\rangle$. The inset shows the time sequence of the measurement. The purple, green, and red areas represent the two-photon transition (TPT) pulses, the clean pulse, and the probe pulse, respectively. In the main plot, circles are the representative data of the difference between the absorption coefficients with and without the second TPT pulse, $\Delta \beta$, as a function of $\Delta t$. The black line is the best fit of an exponential-decay function, which determined the decay rate. In the measurement of these data points, $n_R =$ 0.002 $\mu$m$^{-3}$, and the values of $\Delta$, $\Omega_a$, $\Omega_b$, and $\Omega_{p'}$ were the same as those in shown Fig.~\ref{fig:TPE}.
		}
	\label{fig:Deltabeta}
	\end{figure}
}
\section{Experimental details for determination of parameters}

Before the measurement, we determined the experimental parameters in the order of the coupling Rabi frequency ($\Omega_c$) $\rightarrow$ the intrinsic decoherence rate ($\gamma_0$) $\rightarrow$ the optical depth or OD ($\alpha$). The parameters were confirmed again after the measurement in the reverse order of that before measurement. The method for the determination of experimental parameters is the same as described in Ref.~\cite{OurCommunPhys2021}.

We determined $\Omega_c$ by the Autler-Towns splitting in the EIT spectrum. To measure the frequency seperation between two minima in the spectrum, low OD between 1 and 2 was used. The Autler-Towns splitting was obtained by sweeping the probe field frequency using acousto-optic modulator (AOM). The AOM was not shown in Fig. 1(b) of the main text. The double-path scheme of the AOM can provide the stable input probe power during the frequency sweeping. Degree of asymmetry of the EIT spectrum enabled us to determine the zero coupling detuning condition~\cite{OurPRA2019}.

$\gamma_0$ was determined by the peak transmission of long Gaussian probe pulse at the resonance. The input Gaussian pulse $e^{-1}$ full width was 7~$\mu$s. According to the non-DDI EIT theory, the peak transmission at the resonance condition is given by 
\begin{equation}
T=\rm exp(- \frac{2\alpha\Gamma}{\Omega_c^2}\gamma_0).
\end{equation}
To avoid the dipole-dipole interaction (DDI) induced decoherence effect, we used low OD (15 $\sim$ 20) and probe Rabi frequency $\Omega_p$ ($\sim0.05\Gamma$) for the measurement of $\gamma_0$.

 After $\Omega_c$ and $\gamma_0$ were determined with low OD, we set the OD to the experimental value. We varied the OD by adjusting the repumping laser power during the dark MOT period and the duration of the period. To have the highest OD used in the experiment, we optimized the repumping laser power and the duration. To have a low value of the OD, we changed the repumping power and the duration of the dark MOT from their optimum values.
 
 We measured the delay time $\tau_d$ of the short Gaussian probe pulse to determine the value of OD, i.e., $\alpha$. The $e^{-1}$ full width of the short input Gaussian pulse was 0.66~$\mu$s. According to the non-DDI EIT theory, $\tau_d$ is given by 
\begin{equation}
\tau_d=\frac{\alpha\Gamma}{\Omega_c^2}
\end{equation}


\section{Experimental details for measuring the decay rate of the bright Rydberg state}
\label{subsec:TPT}

\FigSOne

We designed an experiment to determine the coefficients $C$ and $D$ in Eq.~(20). The decay rate, $\Gamma_{24}$, of the atoms in $|2\rangle$ was measured against the atomic density of $|2\rangle$, $n_R$. We employed the two-photon transition (TPT) scheme rather than the EIT scheme to place the population in $|2\rangle$, because the TPT scheme is able to move a large portion of the atoms in $|1\rangle$ to $|2\rangle$. In the measurement of $\Gamma_{24}$, we added the pulses of $\Omega_a$ and $\Omega_b$, which formed the TPT pulse, and the clean pulse as shown in Fig.~\ref{fig:Two-Photon_Transition}(b). The coupling field ($\Omega_c$) in the EIT experiment was utilized as $\Omega_b$, but we changed its frequency to set the one-photon detuning, $\Delta$, to 4.0~$\Gamma$ for the TPT as depicted in Fig.~\ref{fig:Two-Photon_Transition}(a). The frequency of $\Omega_a$ was tuned such that the TPT satisfied the two-photon resonance. A weak and resonant probe pulse, denoted as $\Omega_{p'}$, was used in the measurement. The pulse of $\Omega_{p'}$ had the same optical path as the field of $\Omega_p$ shown in Fig.~1(b). The absorption coefficient of $\Omega_{p'}$ indicated the number of atoms in $|1\rangle$. 

The polarization, propagation direction, and beam size of $\Omega_b$ (or $\Omega_{p'}$) were the same as those of $\Omega_c$ (or $\Omega_p$) in the EIT experiment. The field of $\Omega_a$ was generated by a diode laser, which was injection-locked by the same laser that generated the field of $\Omega_p$ in the EIT experiment. Hence, the two-photon frequency of $\Omega_a$ and $\Omega_b$ was stable. The pulse of $\Omega_a$ was sent to the atom cloud by a PMF after passing through an AOM. The clean pulse was used to remove the population in $|1\rangle$. After passing through the AOM, the clean pulse had a frequency that was resonant to the transition of $|5S_{1/2}, F=2\rangle \rightarrow |5P_{3/2}, F=2\rangle$. The polarization of $\Omega_a$ and the clean pulse were $\sigma_+$ and $\sigma_-$, respectively. There was a separation angle of $0.36^{\circ}$ (or $0.40^{\circ}$) between the propagation directions of $\Omega_a$ (or the clean pulse) and $\Omega_b$. Lens $L_a$ (or $L_c$) was used together with $L_p$ to make $\Omega_a$ (or the clean pulse) a collimated beam with the $e^{-1}$ full width of 3.6 mm (or 4.5 mm). The two beam sizes of $\Omega_a$ and the clean pulse were sufficiently large to cover the entire atom cloud. 

\FigSTwo

To study the efficiency of the TPT pulse, i.e., the simultaneous pulses of $\Omega_a$ and $\Omega_b$, we varied the TPT pulse duration, $\tau_p$, and measured the population or atom number right after the TPT pulse as shown in Fig.~\ref{fig:TPE}. The population in $|1\rangle$ was determined by the absorption coefficient of $\Omega_{p'}$. The timing sequence of $\Omega_a$, $\Omega_b$, $\Omega_{p'}$, and their pulse shapes are depicted in the inset. In the main plot, the circles are the experimental data, and the line is the theoretical prediction of a simple three-level cascade system under the two-photon resonance, i.e., Eqs.~(1)-(6) in the main text are utilized with $\delta = 0$, $\delta_{\rm DDI} = 0$, $\gamma_{\rm DDI} = 0$, $\Omega_p \rightarrow \Omega_a$, and $\Omega_c \rightarrow \Omega_b$. To calculate the prediction, we set $\Delta$, $\Omega_a$, and $\Omega_b$ to the experimental values, and adjusted $\gamma_0$ to match the experimental data. The fair agreement between the data and the prediction verified the effect of the TPT pulse on the population. According to the study shown in Fig.~\ref{fig:TPE}, we decided to use a 1.8~$\mu$s TPT pulse with $\Delta =$ 4.0$\Gamma$ and $\Omega_a = \Omega_b =$ 1.2$\Gamma$ in the measurement of $\Gamma_{24}$ because the transition probability after $\tau_p =$ 1.8~$\mu$s was insensitive to the pulse width. Please note that the reduced population in $|1\rangle$ was used to determine the population in $|2\rangle$, but during the TPT pulse the excitation to the intermediate state $|3\rangle$ was not negligible. Nevertheless, since the population in $|3\rangle$ decayed only to $|1\rangle$ in a rather short time right after the TPT pulse, the excitation of the population to $|3\rangle$ did not affect the determination of the population in $|2\rangle$.

\section{Details on determining of the decay rate from bright to dark Rydberg states}
\label{subsec:Gamma24}

\FigSThree

To obtain the decay rate, $\Gamma_{24}$, of the atoms in $|2\rangle$ as a function of the density, $n_R$, of the atoms in $|2\rangle$, we performed the measurement described in the first paragraph of Subsec.~III B of main text. The time sequence of the measurement is depicted in the inset of Fig.~\ref{fig:Deltabeta}. We kept the time difference between the peak of the clean pulse and the rising edge of the second TPT pulse to about 400~ns, and that between the falling edge of the second TPT pulse and the rising edge of the probe pulse to about 1~$\mu$s. Each of the two TPT pulses had a duration of 1.8~$\mu$s. The TPT pulse was able to move a fixed fraction (about half) of the atoms from the initial state to the final state, i.e., $|1\rangle$ $\rightarrow$ $|2\rangle$ or $|2\rangle$ $\rightarrow$ $|1\rangle$. The residual atoms either were removed from the system or did not affect the measurement of the decay rate. We varied $\Delta t$ and measured the absorption coefficient, $\Delta \beta$, of the weak probe pulse, $\Omega_{p'}$. The representative data of the difference between the values of $\Delta \beta$ with and without the second TPT pulse as a function of $\Delta t$ are plotted in Fig.~\ref{fig:Deltabeta}. Since the atoms in $|1\rangle$ right before the second TPT were all depleted by the clean pulse, $\Delta \beta$ corresponded to the remaining atoms in $|2\rangle$ after a decay time of $\Delta t$.

The data points in Fig.~\ref{fig:Deltabeta} were all taken at $\Delta \alpha =$ 2.9, where $\Delta \alpha$ is defined as the difference of the values of OD before and after the first TPT pulse. The two values of OD were about 5.2 and 2.3 in this case. We were able to estimate the density of the atoms in $|2\rangle$ after the first TPT pulse from $\Delta \alpha$ using the following relation:
\begin{equation}
\label{eq:Deltaalpha}
	\Delta \alpha = \sigma_0 n_1 L = \sigma_0 n_R L,
\end{equation}
where $\sigma_0$ is the absorption cross section of the resonant probe transition from $|5S_{1/2}, F=2, m_F=2\rangle$ to $|5P_{3/2}, F=3, m_F=3\rangle$, and $n_1$ is the density of the atoms in $|1\rangle$ that were moved to $|2\rangle$ by the first TPT pulse. Consequently, $n_1$ was equal to the initial Rydberg-atom density $n_R$, which participated in the decay process of $|2\rangle$. As shown in Fig.~\ref{fig:Deltabeta}, we fitted the experimental data taken at a given $n_R$ with an exponential-decay function. The decay time constant of the best fit gave the value of $\Gamma_{24}$ of the given $n_R$.\\

\section{PREVIOUS STUDIES ON THE POPULATION TRANSFER TO DARK RYDBERG STATES}

Experimental observations of population transfer from one Rydberg state to another, i.e., from a bright Rydberg state to the dark Rydberg state, have been reported in several articles \cite{Martin_2004, Porto_2016,Porto_2017, Hond_2020, Gorshkov_2020,PRA_SR_2007,NJP_SR_2021}. The underlying mechanisms of such transfers can be transitions driven by a microwave field \cite{Martin_2004}, the spontaneous decay induced by black-body radiation and vacuum fluctuations \cite{Porto_2016,Porto_2017,Hond_2020}, the DDI-induced antiblockade excitation and state-exchange collision assisted by radiation trapping \cite{Gorshkov_2020}, and the superradiance induced by black-body radiation (BBR)~\cite{PRA_SR_2007,NJP_SR_2021}. We discuss these articles and compare their results and experimental conditions with our observations and experimental condition in the following paragraphs.

The authors in Ref.~\cite{Martin_2004} measured the linewidth of the transition from $|45D_{5/2}\rangle$ to $|46D_{5/2}\rangle$, while the entire Rydberg population was initially prepared in $|45D_{5/2}\rangle$. As they moved half of the Rydberg population to $|46P_{3/2}\rangle$ by applying a microwave before the linewidth measurement, the measured linewidth was broadened. The authors explained that the DDI between a $|45D_{5/2}\rangle$ atom and a $|46P_{3/2}\rangle$ atom was stronger than that between two $|45D_{5/2}\rangle$ atoms, resulting in the linewidth broadening. The population transfer from the bright ($|45D_{5/2}\rangle$) to dark ($|46P_{3/2}\rangle$) Rydberg state was intentionally driven by a microwave field, and no accumulative DDI effect was reported in Ref.~\cite{Martin_2004}. Since we did not apply any additional field to move the population from the bright to dark Rydberg states, the physical mechanism in this reference is unable to explain the accumulative DDI effect observed in our study.

On the other hand, the population in the dark Rydberg state could be produced via the spontaneous decay from the bright Rydberg state induced by the black-body radiation and vacuum fluctuation. In Ref.~\cite{Porto_2016}, the authors measured the spectrum of the two-photon transition from a ground state to the Rydberg state $|18S_{1/2}\rangle$ with the $^{87}$Rb atoms trapped in a 3D optical lattice. They observed that the measured linewidth was about two orders of magnitude larger than the expected linewidth due to the DDI between two $|18S_{1/2}\rangle$ atoms. Such a large linewidth was explained by the DDI between an atom in $|18S_{1/2}\rangle$ and another in a nearby Rydberg state, $|17P\rangle$ or $|18P\rangle$. The existence of the population in $|17P\rangle$ and $|18P\rangle$ due to the spontaneous decay from $|18S_{1/2}\rangle$ was further verified by Ref.~\cite{Porto_2017}. In Ref.~\cite{Hond_2020}, the authors reported a similar phenomenon with the bright Rydberg state of $|28D_{5/2}\rangle$ and the dark Rydberg states of $|26F_{7/2}\rangle$, $|27F_{7/2}\rangle$, $|29P_{3/2}\rangle$, and $|30P_{3/2}\rangle$. Nevertheless, the spontaneous decay was an one-body process and did not depend on the Rydberg-atom density. As shown in Fig.~6 of main text, the observed decay rate linearly depended on the atomic density of the bright Rydberg state. Furthermore, it was also larger than the spontaneous decay rate of $|32D_{5/2}\rangle$ used in our experiment, which was 2$\pi$$\times$7.9~kHz or 1.3$\times 10^{-3}$$~\Gamma$. Thus, the decay from the bright to dark Rydberg states observed in this study is unable to be explained by the spontaneous decay.

Population transfer from the bright to dark Rydberg states can also be induced by the direct antiblockade excitation \cite{antiblockade_PRL2007, antiblockade_PRL2010} and state-changing Rydberg collisions \cite{state_changing_collision_PRA2015, PRL_stateMixing2008, PRR_stateMixing2020, PRL_stateMixing2004}. In Ref.~\cite{Gorshkov_2020}, the authors drove the Rydberg-EIT transition from a ground state to $|111S_{1/2}\rangle$ (the bright Rydberg state) and detected the ions coming from a number of dark Rydberg states (nearby states other than $|111S_{1/2}\rangle$) after an ionization pulse. Compared with the experiment in our work, the experiment in the reference was carried out under a high atomic density of 5$\times$10$^{12}$~cm$^{-3}$ and in the strong-interaction regime of $(r_B/r_a)^3 > 1$. Under such an atomic density, the authors in Ref.~\cite{Gorshkov_2020} explained that the effect of radiation trapping~\cite{radiationtrapping} was prominent, producing more atoms in the bright Rydberg state. The atoms produced by the radiation trapping had all possible angular momentum angles. Then, the strong-interaction regime enabled the DDI-induced antiblockade excitation and state-changing collisions to generate dark Rydberg atoms from the bright Rydberg atoms. These dark Rydberg atoms produced ions after an ionization pulse, while the ions were clearly not able to come from the bright Rydberg state $|111S_{1/2}\rangle$. The dark Rydberg atom also resulted in the unexpected reduction of the output probe photon number. The authors stated that the reduction is unable to be explained by the prediction of the blockade effect with only the bright Rydberg state but no dark Rydberg states.

A faster population transfer from the bright to dark Rydberg state than the spontaneous decay rate can be explained by the superradiance induced by BBR~\cite{PRA_SR_2007,NJP_SR_2021}. In Ref.~\cite{NJP_SR_2021}, the authors observed the decay of the population from $|nD_{5/2}\rangle$ to $|(n+1)P_{3/2}\rangle$ in laser-cooled caesium Rydberg atoms where $n$ is principal quantum number of 60, 63, and 70. The size of the atom cloud was 550 $\mu$m of diameter. The bright Rydberg state, $|nD_{5/2}\rangle$, is driven by two-photon excitation. $|(n+1)P_{3/2}\rangle$ is one of the dark Rydberg state which is energetically closest state. The population of the bright and dark Rydberg state was measured by the state-selective ionization method with multichannel plate. The authors showed the decay rate from $|nD_{5/2}\rangle$ to $|(n+1)P_{3/2}\rangle$ became faster with larger number of the bright Rydberg state atoms. For example, when they prepared $2.2\times10^4$ number of atoms in $|60D_{5/2}\rangle$, they observed $\sim$2 MHz decay rate which was nearly 660-fold faster decay rate than the decay rate at room temperature. In this case, the effective interaction volume for the cooperative interaction was on the order of $\sim$($\lambda_0$/1000)$^3$ where $\lambda_0$ is transition wavelength of $|60D_{5/2}\rangle\rightarrow|61P_{3/2}\rangle$ transition, 92.9 mm.

